\documentclass{aa}  

\usepackage{graphicx}
\usepackage{txfonts}
\usepackage{subfig}
\usepackage{threeparttable}
\usepackage{amsmath}
\usepackage{mathtools}
\usepackage{xspace}
\usepackage{listings}
\usepackage{color}
\usepackage{placeins}

\definecolor{dkgreen}{rgb}{0,0.6,0}
\definecolor{gray}{rgb}{0.5,0.5,0.5}
\definecolor{mauve}{rgb}{0.58,0,0.82}

\usepackage[usenames]{xcolor}
\usepackage{color}

\newcommand{\rev}[1]{#1}

\newcommand{\Froll}{\ensuremath{F_{\mathrm{roll}}}\xspace}
\newcommand{\vstick}{\ensuremath{v_{\mathrm{stick}}}}
\newcommand{\vbounce}{\ensuremath{v_{\mathrm{b}}}}
\newcommand{\vfrag}{\ensuremath{v_{\mathrm{f}}}}
\newcommand{\vfragz}{\ensuremath{v_{\mathrm{f,0}}}}
\newcommand{\vrms}{\ensuremath{v_{\mathrm{rms}}}}
\newcommand{\amono}{\ensuremath{a_{\mathrm{mono}}}}
\newcommand{\amonoz}{\ensuremath{a_{\mathrm{mono,0}}}}
\newcommand{\mum}{\ensuremath{\mu\mathrm{m}}}
\newcommand{\St}{\ensuremath{\mathrm{St}}}
\newcommand{\rhomat}{\ensuremath{\rho_{\bullet}}}
\newcommand{\afrag}{\ensuremath{a_{\mathrm{f}}}}
\newcommand{\abounce}{\ensuremath{a_{\mathrm{b}}}}
\newcommand{\abouncezero}{\ensuremath{a_{\mathrm{b0}}}}
\newcommand{\akol}{\ensuremath{a_\eta}}
\newcommand{\cs}{\ensuremath{c_{\mathrm{s}}}}
\newcommand{\rhogas}{\ensuremath{\rho_{\mathrm{g}}}}
\newcommand{\sigmagas}{\ensuremath{\Sigma_{\mathrm{g}}}}
\newcommand{\mproton}{\ensuremath{m_{\mathrm{p}}}}
\newcommand{\kboltz}{\ensuremath{k_{\mathrm{B}}}}
\newcommand{\omk}{\ensuremath{\Omega_{\mathrm{K}}}}

\begin{document} 

   \title{The bouncing barrier revisited: Impact on key planet
     formation processes \rev{and observational signatures}}
   \titlerunning{Bouncing barrier and planet formation}
   \author{C. Dominik\inst{1} 
     \and C.P. Dullemond\inst{2}}
   \institute{Anton Pannekoek Institute for Astronomy, University of
     Amsterdam, Science Park 904, 1098 XH Amsterdam, The Netherlands,
     \email{dominik@uva.nl}\\
     \and Institute of Theoretical Astrophysics (ITA), Center for
     Astronomy (ZAH), Ruprecht-Karls-Universität Heidelberg,
     Albert-Ueberle-Str. 2, 69120 Heidelberg, Germany,
     \email{dullemond@uni-heidelberg.de}}

   \date{Received May 15, 2023; accepted May 16, 2023}

% \abstract{}{}{}{}{} 
% 5 {} token are mandatory
 
  \abstract
  {A leading paradigm in planet formation is currently the streaming
    instability and pebble accretion scenario. Notably, dust
    must grow into sizes in a specific regime of Stokes numbers in
    order to make the processes in the scenario viable and sufficiently
    effective. The dust growth models currently in use do not
    implement some of the growth barriers suggested to be relevant in
    the literature.}
  {We investigate if the bouncing barrier, when effective, has
    an impact on the timescales and efficiencies of processes such as the
    streaming instability and pebble accretion as well as on the
    observational appearance of planet-forming disks. }
  {We implemented a formalism for the bouncing barrier into the publicly
    available dust growth model \texttt{DustPy} and ran a series of
    models to understand the impact.}
  {We found that the bouncing barrier has a significant effect on the dust
    evolution in planet-forming disks. In many cases, it reduces the
    size of the typical or largest particles available in the disk; it
    produces a very narrow, almost monodisperse, size distribution; and
    it removes most \mum-sized grains in the process, with an impact on
    scattered light images. It modifies the settling and therefore the
    effectiveness of and timescales for the streaming instability and
    for pebble accretion. {An active bouncing barrier may well
      have observational consequences:  It may reduce the strength of
      the signatures of small particles (e.g., the 10\mum{} silicate
      feature), and it may create additional shadowed regions visible
      in scattered light images.}}
  {{Modeling of planet formation that leans heavily on the streaming
      instability and on pebble accretion should take the bouncing
      barrier into account. The complete removal of small grains in
      our model is not consistent with observations. However, this
      could be resolved by incomplete vertical mixing or some level of
      erosion in collisions.}}


   \maketitle
%
%________________________________________________________________

\section{Introduction}

The formation of planets and planetary systems is one of the key
problems in astronomy today.  It is strongly linked with the question
of the formation and evolution of planet-forming disks and in
particular with the evolution of the dust in these disks. Except for
a subset of \rev{planets that might form} in the outer disk by direct
gravitational collapse \citep{Rice_2022_gravitational_collapse}, all
planet formation is believed to \rev{start} with the growth of dust grains
from their original interstellar or molecular cloud sizes into
planetesimals or even planetary cores \citep{Drazkowska_2022_ppvii}.
The modern paradigm focuses on the formation of so-called pebbles by
direct coagulation of dust in the disk.  It has been well established that
such growth, while relatively quick, does not proceed directly to
large bodies, due to the existence of a number of barriers,
specifically the charge barrier \citep{2011ApJ...731...95O},
the drift barrier
\citep{1972fpp..conf..211W,1977MNRAS.180...57W,2008A&A...480..859B},
the fragmentation barrier
\citep{1997ApJ...480..647D,2005A&A...434..971D}, and the
bouncing barrier \citep{2010A&A...513A..57Z}. The existence
of such barriers at first appears to be bad for planet formation, but
they seem to be consistent with the fact that even after a million years, a
substantial fraction of the solids has not grown beyond the size that
is observable at millimeter wavelengths
\citep{2014prpl.conf..339T}. \rev{The existence of barriers has led} to the creation of a new
class of exciting and effective planet formation models in which a
massive reservoir of particles in a specific size range leads to the
formation \citep[streaming
instability,][]{2007Natur.448.1022J,2007ApJ...662..627J} and further rapid
 growth
\citep[pebble accretion,][]{2010A&A...520A..43O,2012A&A...544A..32L,2016A&A...586A..66V}
of large planetesimals. The available sizes are also relevant for
alternative classes of planetesimal formation models, such as turbulent
concentration \citep{2001ApJ...546..496C,2020ApJ...892..120H}, which
are also expected to grow further through pebble accretion. In this
paper, we explore the role of the bouncing barrier when considering the parameters of the dust population.

The paper is structured as follows. In Sec.~\ref{sec-growth-barriers},
we summarize the information about growth barriers in planet-forming disks
and reintroduce the bouncing barrier. In
Sec.~\ref{sec-physical-background}, we describe the setup used to
compute global models including the bouncing barrier. In
Sec.~\ref{sec:results}, we show the results of the model runs, and in
Sec.~\ref{sec:implications}, we highlight important implications of
this work and discuss the results.  Finally, in
Sec.~\ref{sec:conclusions}, we summarize our conclusions.

%\subsection{Observational constrains}
%\textbf{Observations need to be covered}
%- small grains exist
%- large grains do exist
%- SPHERE \& ALMA
%- retention is currently traps.  Is there another way.
%- 

\section{Growth barriers}
\label{sec-growth-barriers}

The majority of available models for the evolution of dust in disks
includes only the drift barrier and the fragmentation barrier, and most models
ignore other barriers such as the bouncing and charging barriers. The
drift and fragmentation barriers are well understood and
established. {They produce results that are broadly seen as in
  agreement with observations.  They do produce particles in a size
  range that produces sub-millimeter emission, as seen with ALMA, for example ,
  and these particles are also essential for modern planet formation
  theories that include the streaming instability and pebble
  accretion.  Challenges remain, including the spectral indices
  \citep{2014prpl.conf..339T}.  Furthermore, the conditions for the
  streaming instability to occur are difficult to meet and usually
  occur only when restricted in time and space \cite[e.g.,][and references
  therein]{Drazkowska_2022_ppvii}.}

Grains grow by coagulation to increasingly large dust aggregates
until these acquire sizes at which they start to fragment again (the
fragmentation barrier). These fragments re-grow and eventually
re-fragment. A growth-fragmentation cycle becomes established, producing a
rather broad particle size distribution.  That size distribution is
dominated mass-wise by the largest particles, but regarding surface area,
it is dominated by the smallest particles
\citep{2011A&A...525A..11B}. Observations of protoplanetary disks at
optical and near-infrared wavelengths have confirmed the existence of
small, micrometer sized grains, while observations at millimeter
wavelength indicate the presence of large, millimeter sized
grains. The broad grain size distribution produced by the
growth-fragmentation cycle is therefore in agreement with
observational boundary conditions.

The fact that most of the mass in the paradigm ends up in larger
grains forms the basis for the new class of planet formation models
based on "pebbles."  In these models, pebbles are first concentrated
by the streaming instability into self-gravitating clumps, collapse
under their own gravity, and then grow by pebble accretion.

Planet formation models that rely on radial drift of vast amounts of
pebbles (i.e., large dust aggregates) from the outer into the
inner disk regions, the so-called pebble accretion models
\citep{2010A&A...520A..43O, 2012A&A...544A..32L}, require that the
grains grow to sufficiently large sizes in order to begin drifting
efficiently. The fragmentation barrier, especially when assuming that
the critical collision velocity for fragmentation (the
"fragmentation velocity") is on the high end of the established
range of values (i.e., at 10 m/s) produces pebbles that are large
enough to enable the required large influx of pebbles into
the planet-forming regions of the disk.

Given the success of these dust coagulation models with only the drift
and fragmentation barriers, it seems to be forgivable to ignore
additional barriers. The consequence of this limitation is that the
leading models using the streaming instability to form planetesimals
and the process of pebble accretion to further grow the largest of
these planetesimals into planetary cores
\citep[e.g.,][]{izidoro_formation_2021} make use of pebbles with the
sizes given by these boundaries. However, if there exist additional
barriers that stop growth at somewhat smaller sizes, this will
have significant impact on the occurrence and strength of the
streaming instability as well as on the efficiency of pebble
accretion.  It may have the effect of extending the duration of dust retention in disks
and slowing down planetary growth.

The existence of the charging barrier and its significant consequences
has already been considered and discussed in detail in a series of
papers
\citep{okuzumi_electric_2009,okuzumi_electrostatic_2011-1,okuzumi_electrostatic_2011}. In
this paper we focus on a re-evaluation of the bouncing barrier
and compute the evolution of global disk models under the presence of
an active bouncing barrier.
The expectation that there should be a regime in collision velocity
where aggregates should be bouncing instead of sticking can already be
derived from energetic considerations
\citep{youdin_obstacles_2004}.
%It is also of course a well-observed
%fact in collisions of solid (non-aggregate) particles \citep{blum_dust_2018,kruss_growing_2017,brisset_low-velocity_2017,kruss_failed_2016,weidling_free_2015,kelling_experimental_2014,guttler_outcome_2010}.

While the first numerical models of soft aggregate collisions
\citep{dominik_physics_1997,wada_numerical_2008,paszun_collisional_2009}
did not show bouncing for fluffy aggregates, it was later shown that
more compact aggregates,
\citep{wada_rebound_2011,schrapler_physics_2012,seizinger_bouncing_2013,weidling_free_2015}
or aggregates in which contacts had been strengthened or stiffened by
sintering \citep{sirono_collisions_2017}, do show bouncing behavior.

Laboratory experiments have shown without doubt that relatively
compact aggregates, when collided at velocities beyond the sticking
threshold, do readily rebound
\citep{hill_collisions_2015,brisset_low-velocity_2017}; this includes
aggregates made of icy particles \citep{gundlach_stickiness_2015}.
Another series of experiments showed that a population of grains can
first grow and then be stalled by the onset of bouncing
\citep{kelling_experimental_2014,kruss_failed_2016}.

When a growth process becomes trapped and stopped by the bouncing
barrier, it might be possible to rely on effects of a broad size
and/or velocity distribution to allow for a small number of particles to eventually 
break through and continue to grow into larger bodies
\citep{windmark_breaking_2012,windmark_planetesimal_2012,kruss_growing_2017,booth_breakthrough_2018}.

A small number of coagulation computations including the bouncing
barrier exist
\citep{guttler_outcome_2010,windmark_planetesimal_2012,drazkowska_planetesimal_2013,zsom_outcome_2010,xiang_detailed_2020,2023A&A...670L...5S}.
But by and large, the process has not been taken into account by the
family of models that underlay modern global planet formation
models. A global model has also been presented by
\cite{2016ApJ...818..200E,2022ApJ...936...42E,2022ApJ...936...40E}. It
implements an efficient moment method for the growth of small
particles below the fragmentation barrier and a histogram
representation of migrator particles beyond the fragmentation barrier,
including a treatment of developing of porosity as well as transport
of volatiles. In this work, we focus on the effect of introducing bouncing in
the context of its implications for planet formation models and disk
observations.

\section{Base model and implementation}
\label{sec-physical-background}

As discussed above, bouncing behavior has been observed in a number of
experimental setups, and it has been reproduced in numerical models of the
mechanical behavior of dust aggregates. We therefore take bouncing as
a given fact for the study presented in this paper. However, we
always compare the results in this work with an equivalent model without bouncing
in order to highlight the differences that occur.

We used the \texttt{DustPy} \citep{stammler_dustpy_2022} modeling
tool. The \texttt{DustPy} tool is a Python package used to simulate dust evolution
in protoplanetary disks. It solves {radial} gas and
dust transport including viscous advection and diffusion as well as
collisional growth of dust particles in detail. {In the direction
  normal to the disk midplane, an equilibrium between vertical
  turbulent mixing and dust settling is assumed to be valid and to
  readjust instantly. In our runs, we assumed that the $\alpha$
  parameter for turbulence is the same value for radial and vertical
  mixing and viscous accretion.} The basic model implements the
well-known standard models of
\citet{birnstiel_gas-_2010}. The \texttt{DustPy} tool is written in a modular
way so that it is easy to extend it and to implement bouncing for the
present study.  Technically, we wrote a routine that computes the
sticking probabilities, taking into account the possibility of bouncing
occurring in a collision, and then we assigned this routine to the
\texttt{sim.dust.p.stick.updater} variable, which then calls on this
routine instead of the built-in routine. For details, see the Appendix
\ref{app-code-snippets}.
%We made the implementation
%available in a github repository that can be found at
%\texttt{github.com/XXXXXXXXXXX}.

\subsection{Implementation}
\label{sec:implementation}

We followed the \cite{birnstiel_gas-_2010} basic model setup by
assuming that grains are spherical and compact and that the
combination of two such particles after a collision into a new,
larger grain is still compact and works with conservation of mass and
volume. In other words, we kept the density of particles constant at
$\rho_{\bullet}=1.67\;\mathrm{g/cm^2}$. This is obviously a
significant simplification since we do know, at least initially, that
the growth of particles produces very low-density fractal structures
\citep{dominik_physics_1997,Krause_Blum_2004_fractal}. However, for
the current paper, we focus strictly on the effect of the bouncing
barrier, so we leave a better treatment of porosity to future studies.

We also chose a simple implementation for the bouncing barrier.
This implementation is based on laboratory
experiments and theoretical scaling laws. The default model in \texttt{DustPy}
has two possible outcomes of a collision, namely, sticking and fragmentation.
We assumed three regimes in the collisional space between two
particles.  For low velocities, we assumed that the particles
simply stick together and form a new particle with the combined mass
and volume.  When the relative velocity in a collision exceeds a
critical velocity $\vbounce(m_1m_2)$ for bouncing, we assumed that
the particles bounce and preserve their identity, mass, and volume.
The critical velocity depends on the masses of the particles
having the collision.  At still higher velocities, we defined
another critical velocity: the fragmentation velocity
$\vfrag(m_1,m_2)$.  We then defined probabilities for sticking,
bouncing, and fragmentation as a function of the masses and relative
velocity of any particle pair.  Since the relative velocities are
not strict functions of the masses, we folded the probabilities
with Maxwellian distributions over the relative velocity.  This gave a
better result and is also numerically favorable since the transitions
between the different regimes are softened in a natural way.

We started by assuming a fragmentation velocity
\begin{equation}
\label{eq:2}
\vfragz=100\mathrm{\,cm\,s^{-1}} \quad
\end{equation}
for particles that are aggregates made of monomers with a radius
$\amonoz=1\,\mum$.  The dependence of the fragmentation speed on the
size of the monomer has not been properly and consistently measured
over a significant range of monomer radii. Here, we applied scalings that
are derived from \cite{dominik_physics_1997}.  We assumed that the
collision energy required to break up an aggregate is given by the
energy to break up a single contact in the aggregate times the number
of contacts in the aggregate times a scaling factor (found to be
$\sim10$ by \cite{dominik_physics_1997}):
$E_{\mathrm{frag}}=10 n_{\mathrm{contacts}}E_{\mathrm{break}}$. We
then followed the scalings of those factors with the monomer radius
\amono.  For a given mass of the aggregate, the number of monomers in
the aggregates scales as $\propto\amono^{-3}$, and the number of
contacts then scales in the same way. The energy to break a contact
scales as $\propto \amono^{4/3}$, which can be derived from the
formulas in \cite{chokshi_dust_1993} and
\cite{dominik_physics_1997}. The same result can be found from the
work of \cite{Krijt_2014_rolling}, using that the breakup energy is
given by the rolling force times the monomer radius. The energy for a
fragmenting collision therefore scales as $\propto\amono^{-5/3}$.  In
this way, we
finally arrived at the result that \rev{we may scale} the fragmentation
velocity with
\begin{equation}
\label{eq:vfragscale}
\vfrag = \vfragz \left(\frac{\amono}{\amonoz}\right)^{-5/6} \quad .
\end{equation}

\begin{figure}
  \includegraphics[width=.45\textwidth]{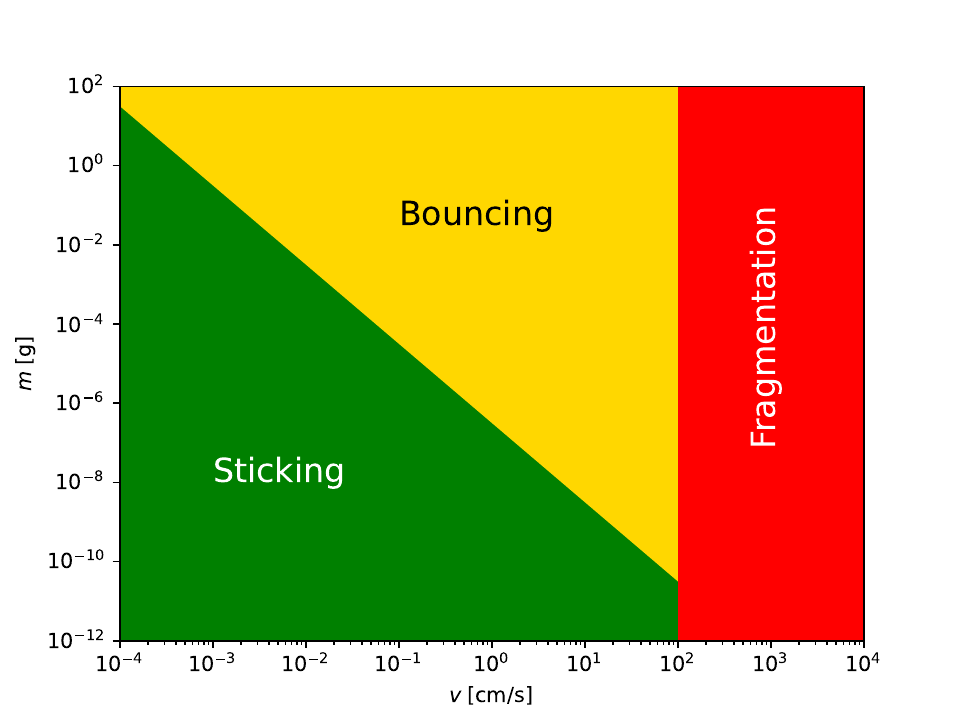}
  \caption{\label{fig-stick-bounce-frag} Sticking-bouncing-fragmentation
    diagram showing in which parts of the parameter space equal-mass collisions
    lead to one of the three outcomes. This figure is the
    direct equivalent of Fig.~11 of \citet{guttler_outcome_2010}, with
    $\Froll=10^{-4}$ dyne, $\amono=1\,\mu$m, and $\vfrag=100$ cm/s.}
\end{figure}

For the critical velocity for bouncing, we started by looking at the
hit-and-stick velocity for dust aggregates
\citep{dominik_physics_1997,guttler_outcome_2010}. This velocity is
defined by assuming that the relative velocity is too small
to significantly deform even a fractal aggregate.  This velocity is given
by
\begin{equation}
\label{eq:vstick}
\vstick(m_1,m_2) = \sqrt{5\frac{\pi \amono \Froll}{m_{\mu}}},
\end{equation}
where \Froll is the force needed to roll a monomer in contact over
another monomer and $m_{\mu}=\frac{m_1m_2}{m_1+m_2}$ is the reduced mass
in the collision of two particles with masses $m_1$ and $m_2$,
respectively. We took $\Froll=10^{-4}$ dyne, which is roughly the
value determined experimentally by \citet{1999PhRvL..83.3328H}.
Once an aggregate particle was sufficiently compacted so
that rearrangement of monomer-monomer contacts is no longer possible
without fragmenting the particle, we assumed that these particles
would bounce.  For our simple implementation of the bouncing barrier, we
therefore assumed bouncing to occur when the relative velocity between
two particles $\Delta v$ is between
\begin{equation}
\vbounce \le \Delta v<\vfrag,
\end{equation}
where
\begin{equation}\label{eq-vbounce}
\vbounce = \left\{\begin{matrix}
&\vstick(m_1,m_2) &\quad &\hbox{when} & \quad & \vstick(m_1,m_2)\le \vfrag \\
&\vfrag &\quad &\hbox{when} & \quad & \vstick(m_1,m_2)> \vfrag.
\end{matrix}\right.
\end{equation}
This led to the well-known stick-bounce-fragment diagram of the type
introduced by \citet{guttler_outcome_2010}, shown here in Fig.~\ref{fig-stick-bounce-frag}.

\begin{figure}
  \includegraphics[width=.45\textwidth]{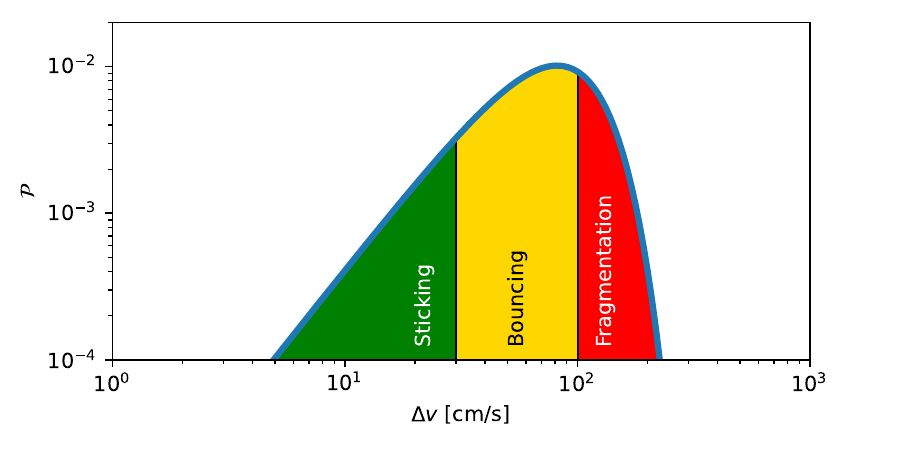}
  \caption{\label{fig-maxboltz}Assumed Maxwell-Boltzmann-like distribution
    function for the relative velocities between colliding dust particles
    and the three regimes of collisional outcomes. This figure is the
    direct equivalent of Fig.~6-Left of \citet{2022ApJ...935...35S}, with
    $\vbounce=30$ cm/s and $\vfrag=100$ cm/s.}
\end{figure}

Following \citet{windmark_breaking_2012} and \citet{2022ApJ...935...35S}, we computed the probability of sticking,
bouncing, and fragmentation for any pair of dust particle masses $(m_1,m_2)$ by
assuming that the relative velocities of these dust particles obey a
Maxwell-Boltzmann distribution defined as 
\begin{equation}\label{eq-maxwell-boltzmann}
  {\cal P}_{\mathrm{MB}}(\Delta v) = 3\sqrt{\frac{6}{\pi}}\,\frac{\Delta v^2}{\vrms^3}
  \,\exp\left(-\frac{3}{2}\frac{\Delta v^2}{\vrms^2}\right),
\end{equation}
where $\vrms$ is the root-mean-square of the relative
velocities. The probability of a collision between these two dust
particles with a relative velocity greater than a given value $v$ is then
\begin{equation}\label{eq-cumul-maxbolz}
  P(\Delta v>v) = \int_v^\infty {\cal P}_{\mathrm{MB}}(\Delta v) d\Delta v
  = \left(\frac{3}{2}\frac{v^2}{\vrms^2}+1\right)
  \exp\left[-\frac{3}{2}\frac{v^2}{\vrms^2}\right]
\end{equation}
\citep[see Eq.~10 of][]{2022ApJ...935...35S}. By setting $v=\vfrag$ in
Eq.~(\ref{eq-cumul-maxbolz}), we obtained the probability that a collision leads to
fragmentation. By setting $v=\vbounce$ in Eq.~(\ref{eq-cumul-maxbolz}), we obtained
one minus the probability that a collision leads to sticking. This is
illustrated in Fig.~\ref{fig-maxboltz}.

Without bouncing, these two probabilities add up to unity. However,
with bouncing, they add up to a value below unity. The remainder is
the probability of bouncing. Since bouncing does not change the mass
distribution of the dust, the mere fact that the probabilities of
sticking and fragmentation do not add up to one implies that the
bouncing barrier is implemented. Bouncing is simply a reduction of the
effective collision rate. In the Appendix \ref{app-code-snippets}, we
give Python code snippets that we used to implement this bouncing
barrier model in \texttt{DustPy}.

\subsection{Estimation of the bouncing barrier}
\label{sec:estimate-bb}
\begin{figure*}
  \begin{center}
    \includegraphics[width=\textwidth]{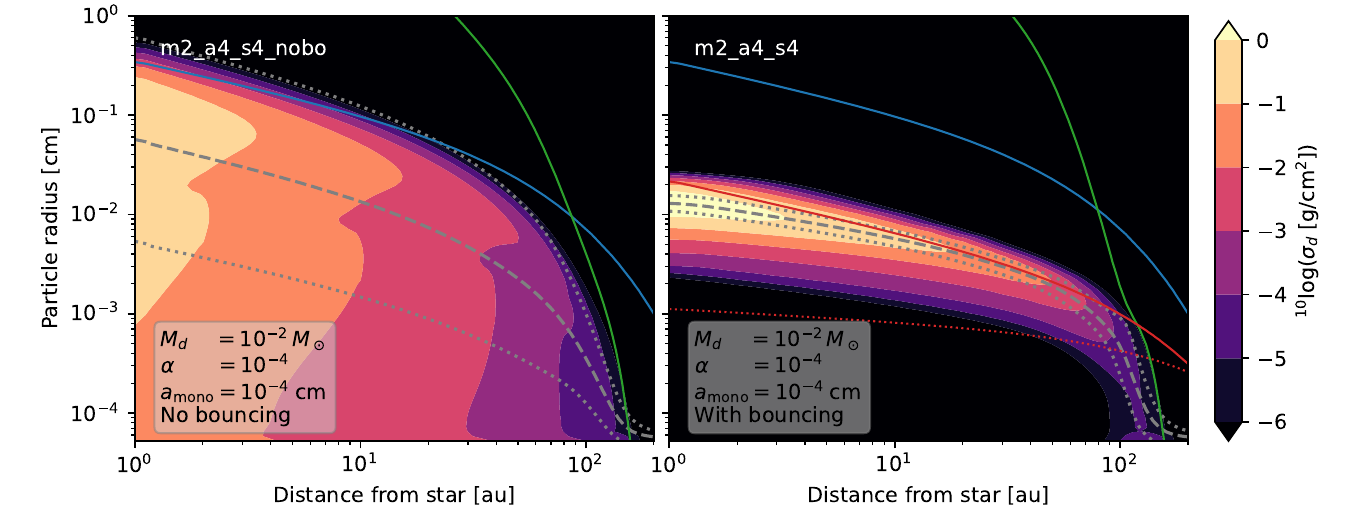}
  \end{center}
  \caption{\label{fig-fidu-sigma}Dust size distribution for the fiducial
    model without the bouncing barrier \texttt{m2\_a4\_s4\_nobo} (left) and with
    bouncing barrier \texttt{m2\_a4\_s4} (right). The blue line is the
    fragmentation barrier, the green line is the radial drift barrier, and the
    red line is the bouncing barrier, all computed assuming turbulence to be the
    dominant driver of relative velocities. {The dotted red line is the
      bouncing barrier if Eqs.~(\ref{eq-akol},\ref{eq-abounce-with-akol})
      are not taken into account.} The dashed gray line is the mean
    grain size, and the dotted lines are plus or minus one standard deviation in
    log space (see Appendix \ref{app-mean-and-width}). The color scale denotes
    the surface density of the dust $\sigma_{\mathrm{d}}(r,m)=m\,\Sigma_{\mathrm{d}}(r,m)$ in
    units of grams per square cm$^{2}$ per $\ln(m)$.}
\end{figure*}

Under the assumption that turbulence is the dominant cause of relative
velocities between dust grains, \citet{2012A&A...539A.148B} formulated analytic
estimates of the growth mass limits due to fragmentation and radial drift and
showed them to be good predictors for the upper envelope of the grain size
distribution. The largest Stokes number the grains acquire is given by
\begin{equation}
  \label{eq:stokes-fragmentation}
\St_{\mathrm{f}} = \frac{1}{3}\frac{\vfrag^2}{\alpha \cs^2},
\end{equation}
where $\cs$ is the isothermal sound speed of the gas and $\alpha$ is the usual
turbulence parameter. According to
\citet{birnstiel_gas-_2010}, the {radius $a$ and the Stokes number $\St$} are related via
\begin{equation}
  \label{eq:stokes-number}
\St = \frac{\rhomat a}{\sigmagas}\,\frac{\pi}{2},
\end{equation}
where $\sigmagas$ is the surface density of the gas in the disk {and $\rhomat$
is the material density of the dust grain, assuming a spherical grain geometry and
with porosity included}. {Solving
Eq.~(\ref{eq:stokes-fragmentation}) for $a$, we obtained the grain size
limit due to fragmentation:}
\begin{equation}
  \label{eq:afrag}
\afrag = \frac{2}{3\pi}\frac{\sigmagas\vfrag^2}{\alpha\rhomat\cs^2}.
\end{equation}
For the bouncing barrier, we suggest a similar formula:
\begin{equation}\label{eq-stokes-bounce}
\St_{\mathrm{b0}} = \frac{1}{3}\frac{\vbounce^2}{\alpha \cs^2},
\end{equation}
where $\vbounce(m)=\vbounce(m,m)$ (i.e., we assume equal-sized
collisions). This is a pessimistic estimate because in a collision,
strongly unequal-sized particles are more likely to stick than
particles of equal size, as one can see from
Eq.~(\ref{eq:vstick}). However, \rev{as shown below}, once the bouncing
barrier becomes dominant, the dust distribution becomes rather narrow,
meaning that collisions of similar-sized particles become the most
common. We note that we added a zero to $\St_{\mathrm{b0}}$ because as
we show later {(Eqs.~\ref{eq-akol},\ref{eq-abounce-with-akol} and the
  red dotted line in Fig.~\ref{fig-fidu-sigma})}, this estimate has to
be modified due to the Kolmogorov cutoff scale. However, we initially
ignore this effect.  Equation ~(\ref{eq-stokes-bounce}) is an implicit
equation for the bouncing barrier {limit on the grain size} because
both the Stokes number $\St$ and the bouncing velocity $\vbounce$
depend on the {grain} mass $m$.  The {dust grain} mass $m$ and radius
$a$ are related via $m=4\pi\rhomat a^3/3$, allowing us to express
$\vbounce(m)$ in terms of $a$. Then solving
Eq.~(\ref{eq-stokes-bounce}) for $a$, we obtained the grain size limit
due to bouncing:
\begin{equation}
  \label{eq:abouncezero}
\abouncezero = \left(\frac{5}{\pi}\frac{\sigmagas \amono \Froll}{\alpha \cs^2 \rhomat^2}\right)^{1/4}.
\end{equation}
{Apparent from this equation is the fact that the bouncing barrier is rather
  insensitive to the disk parameters, such as $\alpha$ and $\sigmagas$, due
  to the one-fourth power dependency. When $\alpha$ is increased by a factor of ten, the
  bouncing barrier grain size $\abouncezero$ decreases by only 15\%. The
  fragmentation barrier (Eq.~\ref{eq:afrag}) does not have this one-fourth power
  dependency. Thus, when $\alpha$ is increased by a factor of ten, the
  fragmentation barrier grain size $\afrag$ decreases by a factor of ten. As
  we later show, this has the consequence that for a high $\alpha$, the
  fragmentation becomes important, while for a low $\alpha$, the grain growth
  is stopped by the bouncing barrier before reaching the fragmentation barrier.
  We also show that this has interesting consequences for models of
  chondrule formation (Section \ref{sec-chondrule-formation}).}

However, it turns out that this estimate of the grain size of the bouncing
barrier is often too small because the Kolmogorov turbulence cascade does not
continue to small enough eddies. This problem rarely happens when computing the
fragmentation barrier because that occurs at rather large velocities.  Yet it
becomes important for the bouncing barrier. In the turbulence model of
\citet{2007A&A...466..413O}, the Kolmogorov scale is computed as follows: The
gas of the disk is assumed to be dominated by H$_2$ molecules, which have a
mutual collisional cross section of $\sigma_{\mathrm{H}_2}=2\times
10^{-15}\,\mathrm{cm}^2$. The mean molecular weight is assumed to be $\bar
m=2.3\,\mproton$, where $\mproton$ is the proton mass. This leads to a mean free path of
the gas particles of $\lambda_{\mathrm{mfp}}=\bar
m/\rhogas\sigma_{\mathrm{H}_2}$, where $\rhogas$ is the gas volume density.  The
thermal velocity of the gas particles is $v_{\mathrm{th}}=\sqrt{8\kboltz{}T/\pi \bar
  m}=\sqrt{8/\pi}\,\cs$, where $\kboltz{}$ is the Boltzmann constant and $T$ is the
temperature of the gas.  This leads to a molecular viscosity of
$\nu_{\mathrm{mol}}=v_{\mathrm{th}}\lambda_{\mathrm{mfp}}/2$. The Reynolds
number of the largest eddies is Re = $\alpha \cs h_p/\nu_{\mathrm{mol}}$, where
$h_p=\cs/\omk$ is the scale height of the disk, with $\omk$ the Kepler
frequency. The turnover timescale of the eddies at the Kolmogorov scale becomes
$t_\eta=(\omk\sqrt{\mathrm{Re}})^{-1}$. The grain size coupling to these
smallest eddies is then
\begin{equation}\label{eq-akol}
\akol = \frac{\rhogas \sqrt{8/\pi}\cs t_\eta}{\rhomat} = \frac{\rhogas
  \cs}{\rhomat\omk}\sqrt{\frac{8}{\pi\,\mathrm{Re}}}\quad .
\end{equation}
The grain size of the bouncing barrier is then
\begin{equation}\label{eq-abounce-with-akol}
\abounce = \max(\abouncezero,\akol) \quad .
\end{equation}
{In Section \ref{sec:results}, in Fig.~\ref{fig-fidu-sigma}, the red
  solid an dotted lines show the difference this makes for our fiducial model.} 

We note that \citet{2021ApJ...917...82G} propose a modification of the turbulent
stirring model of \citet{2007A&A...466..413O}, which may better describe the
turbulence produced by the magneto-rotational instability. In their model,
the Kolmogorov cutoff scale is at a smaller Stokes number, and the relative
velocities between particles are generally larger. This would push the
bouncing barrier toward smaller aggregate sizes. 

\begin{table}[b]
\begin{center}\small
\caption{\label{table-params}\small Model parameters.}
\begin{tabular}{l|l|cccc}
  Series             & Model name                         & $M_{\mathrm{disk}}\; [M_{\odot}]$ & $\alpha$            & $s\; [\mathrm{cm}]$ & BB           \\\hline
  $\alpha$           & \texttt{m2\_a3\_s4\_nobo}          & $10^{-2}$                         & $\mathbf{10^{-3}}$  & $10^{-4}$           & no           \\
  \textbf{fiducial}  & \texttt{\textbf{m2\_a4\_s4\_nobo}} & $\mathbf{10^{-2}}$                & $\mathbf{10^{-4}}$  & $\mathbf{10^{-4}}$  & \textbf{no}  \\
  $\alpha$           & \texttt{m2\_a5\_s4\_nobo}          & $10^{-2}$                         & $\mathbf{10^{-5}}$  & $10^{-4}$           & no           \\
  $\alpha$           & \texttt{m2\_a6\_s4\_nobo}          & $10^{-2}$                         & $\mathbf{10^{-6}}$  & $10^{-4}$           & no           \\  
  misc               & \texttt{m1\_a4\_s4\_nobo}          & $\mathbf{10^{-1}}$                & $10^{-4}$           & $10^{-4}$           & no           \\
  misc               & \texttt{m2\_a4\_s5\_nobo}          & $10^{-2}$                         & $10^{-4}$           & $\mathbf{10^{-5}}$  & no           \\[2mm]
  $\alpha$           & \texttt{m2\_a3\_s4}                & $10^{-2}$                         & $\mathbf{10^{-3}}$  & $10^{-4}$           & yes          \\
  \textbf{fiducial}  & \texttt{\textbf{m2\_a4\_s4}}       & $\mathbf{10^{-2}}$                & $\mathbf{10^{-4}}$  & $\mathbf{10^{-4}}$  & \textbf{yes} \\
  $\alpha$           & \texttt{m2\_a5\_s4}                & $10^{-2}$                         & $\mathbf{10^{-5}}$  & $10^{-4}$           & yes          \\
  $\alpha$           & \texttt{m2\_a6\_s4}                & $10^{-2}$                         & $\mathbf{10^{-6}}$  & $10^{-4}$           & yes          \\
  misc               & \texttt{m1\_a4\_s4}                & $\mathbf{10^{-1}}$                & $10^{-4}$           & $10^{-4}$           & yes          \\
  misc               & \texttt{m2\_a4\_s5}                & $10^{-2}$                         & $10^{-4}$           & $\mathbf{10^{-5}}$  & yes  
\end{tabular}
\end{center}
\small Note: The name (column 2) was
  constructed from minus the exponents of the values in columns 3, 4,
  and 5, respectively. Column 3 is the disk gas+dust mass in units of
  solar mass. Column 4 is the turbulence parameter $\alpha$. Column 5
  is the monomer size $s$ in units of centimeters. Column 6 lists whether
  the bouncing barrier was included in the model or not. The full rows
  for the models \texttt{\textbf{m2\_a4\_s4\_nobo}} and
  \texttt{\textbf{m2\_a4\_s4}} are in bold to indicate that
  these are the fiducial models. For the other models, we
  have made bold the parameter that deviates from the
  fiducial model. Column 1 shows to which series the model belongs
  (for the discussion, we refer to Sec.~\ref{sec:results}).
\end{table}

\begin{figure}
  \includegraphics[width=.45\textwidth]{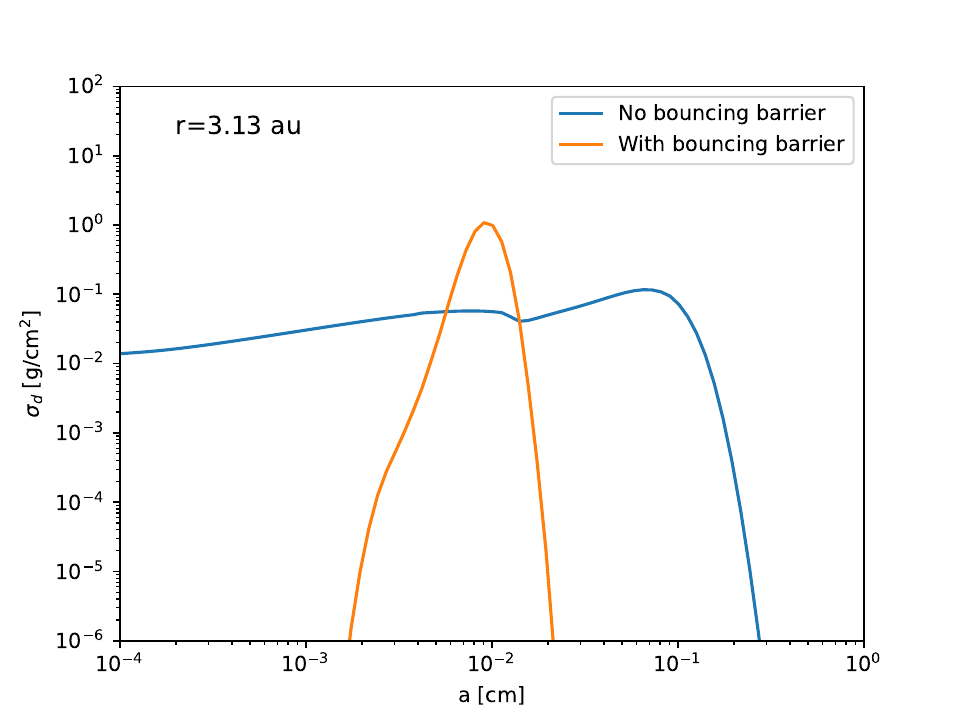}
  \caption{\label{fig-fidu-slice-sigma}As in Fig.~\ref{fig-fidu-sigma} but now with a
  slice at $r=3.13$ au.}
\end{figure}

\section{Results}
\label{sec:results}
\rev{In this section, we describe the parameters used in our computations.
We then show the results.}

\subsection{Model parameters}
As an initial condition, we set up a protoplanetary disk with the usual power law plus cutoff shape,
\begin{equation}\label{eq-gas-surface-density}
\sigmagas(r) = \Sigma_0\,\left(\frac{r}{r_0}\right)^{-1}\exp\left(-\frac{r}{60\,\mathrm{au}}\right),
\end{equation}
around a solar-type star. The initial dust-to-gas ratio is 0.01. We chose
$\Sigma_0$ and $r_0$ such that the mass of the disk is a given value
$M_{\mathrm{disk}}$, where $M_{\mathrm{disk}}$ is a parameter we 
varied. Other parameters we varied are the turbulent strength $\alpha$ and the
monomer radius $s$. The radial $r$ grid is between
$r_{\mathrm{in}}=1\,\mathrm{au}$ and $r_{\mathrm{out}}=10^3\,\mathrm{au}$ with
$N_r=$100 logarithmically spaced radial bins. The coagulation was computed on a mass
grid between $m_{\mathrm{min}}=10^{-12}\,\mathrm{g}$ and
$m_{\mathrm{min}}=10^{5}\,\mathrm{g}$ with 120 logarithmically spaced
bins\footnote{{We tested higher grain size resolutions and found 120 bins
to be sufficient for our application. We also ran tests with different time
steps.  {Due to the implicit integration of the equations, \texttt{DustPy}
works well even for a moderately boosted time step.}
For the calculations shown in this paper, the boost factor was
set to ten.  We tested runs with this factor set to one and 
100.  The differences between the runs with factors one and ten are
insignificant. Runs did become numerically unstable with a boost
factor of 100.  For details see the Appendix.~\ref{app-resolution}.}}. The
material density was taken to be
$\rho_{\bullet}=1.67\,\mathrm{g}\,\mathrm{cm^{-3}}$. The fragmentation velocity
was set at 100 cm/s and scaled only when the monomersize was varied,
using Eq.~\eqref{eq:vfragscale}. The fragment distribution function is the standard power law
$f(m)\propto m^{-11/6}$, which is \rev{identical to the
Mathis-Rumple-Nordsieck \citep[MRN,][]{1977ApJ...217..425M} size distribution.}  The
models were run for $10^6$ years, and we show the results for the final time
frame at $t=10^6$ years.

The parameter space spanned up by $M_{\mathrm{disk}}$, $\alpha$, and
$s$ was explored in steps of ten. The parameter values are listed
in Table~\ref{table-params}.

\subsection{Fiducial model}
\label{sec:fiducial}
To discuss our results, we chose as our fiducial model the model with
$M_{\mathrm{disk}}=10^{-2}M_{\odot}$, $\amono=10^{-4}\,\mathrm{cm}$ and
$\alpha=10^{-4}$ (model \texttt{m2\_a4\_s4}). For comparison, we also discuss the
results of model {\sf m2\_a4\_s4\_nobo}, which is the same model but without
the bouncing barrier.

To analyze the results of the models, we computed a number of
quantities at each radius $r$ from the local dust size distribution:
the mean and the width of the distribution, the vertical optical
depths, the mean degree of settling, the number of collisions per
megayear, and the mean collisional velocity. Furthermore, images at two
wavelengths were computed. The details of how these quantities and
images were computed are described in the Appendix \ref{app-diagnostics}.

The comparison between the non-bouncing and the bouncing versions of
the fiducial model are shown in Fig.~\ref{fig-fidu-sigma}. The
original \texttt{DustPy} model without the bouncing barrier is shown on the left, and
the new model with the bouncing barrier is shown on the right. A slice of the
same data at $r=3.13$ au is shown in Fig.~\ref{fig-fidu-slice-sigma}
in a log-log figure.

Two main effects of the bouncing barrier are immediately obvious from
these figures. One is that the grains stop growing at a smaller grain
size, in this case up to a factor of ten in radius or 1000 in mass. The
other is that the grain size distribution becomes very narrow, almost
monodisperse. This is very different from the almost flat
distribution created by the growth-fragmentation steady state in the
model without bouncing. At the outer edge of the disk, the drift barrier
still plays a role. Inward of 100\,au, the bouncing barrier determines the grain
size distribution. \rev{We 
anticipate that this may have consequences for the growth speed of}
the streaming instability and for the accretion rate and efficiency of
pebble accretion. We discuss these consequences in more detail in
Sec.~\ref{sec:si-pf-pa}.

\begin{figure*}
  \includegraphics[width=\textwidth]{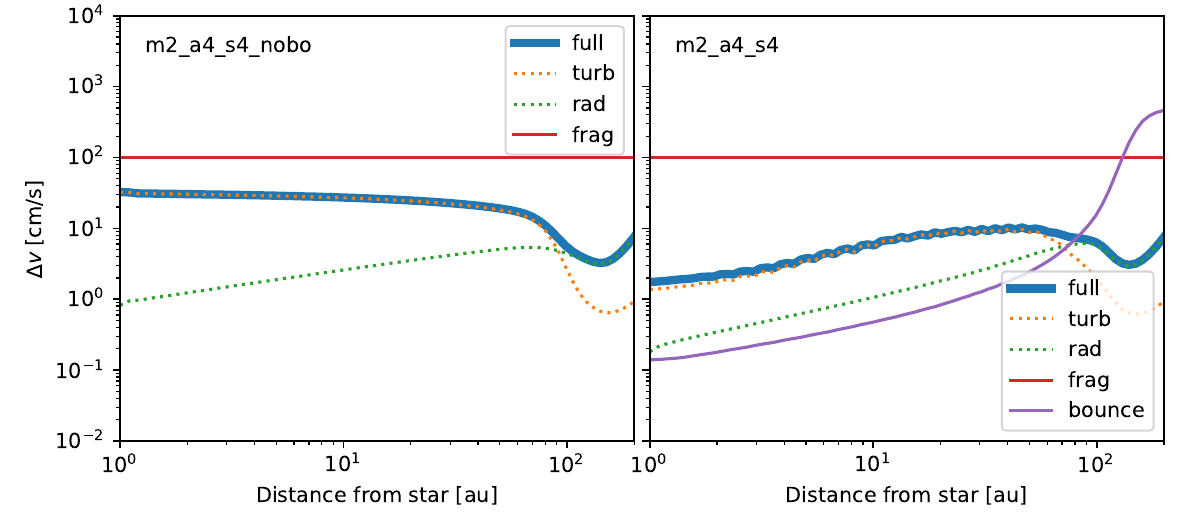}
  \caption{\label{fig-fidu-collision-velocities}Mean collision velocities in the
    fiducial model. Left: Without bouncing. Right: With bouncing.}
\end{figure*}

\begin{figure}[b!]
  \includegraphics[width=0.5\textwidth]{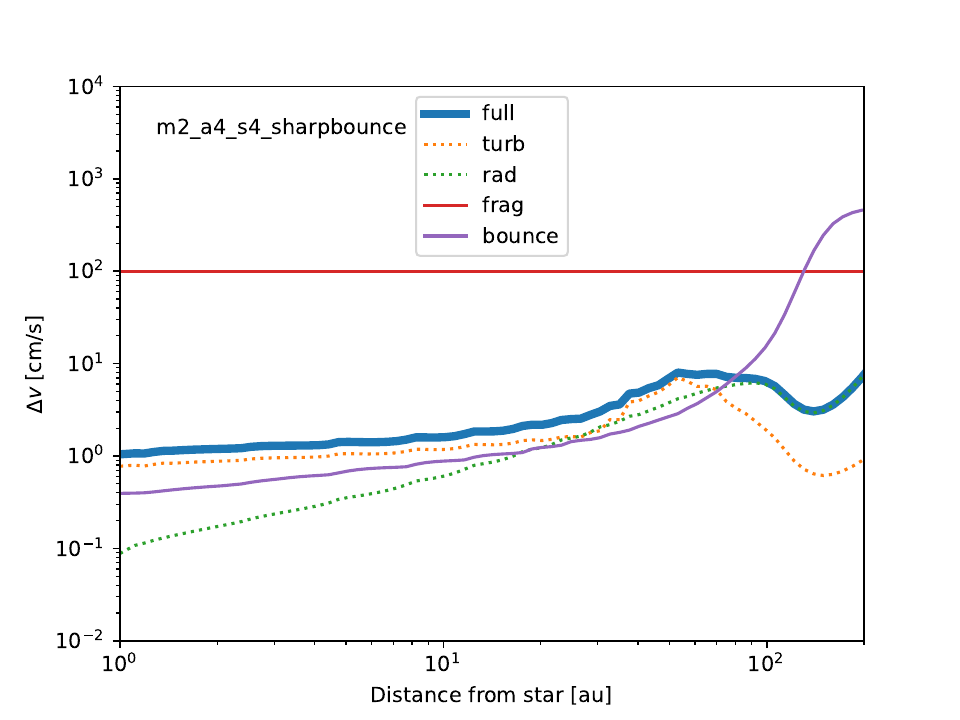}
  \caption{\label{fig-fidu-sharpbounce}As in Fig.~\ref{fig-fidu-collision-velocities}-right
    but this time using the mean collision velocities instead of the Maxwell-Boltzmann distribution for the relative velocities.}
\end{figure}

\begin{figure*}
  \includegraphics[width=\textwidth]{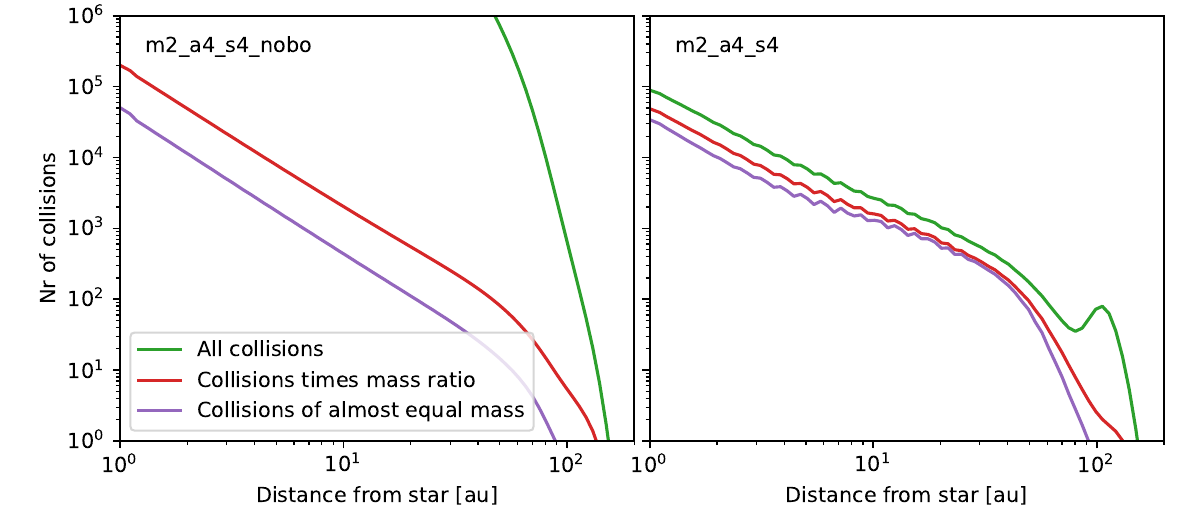}
  \caption{\label{fig-fidu-nr-of-collisions}Mean number of collisions that a
    dust particle experiences over the course of $10^6$ years in the fiducial
    model.  Left: Without bouncing. Right: With bouncing.}
\end{figure*}

\begin{figure*}
  \includegraphics[width=\textwidth]{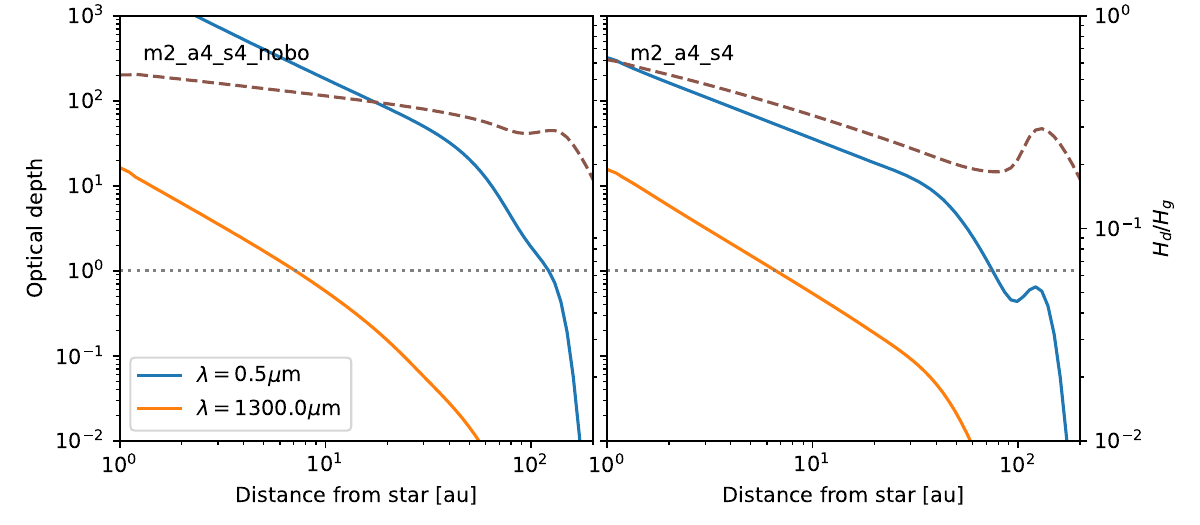}
  \caption{\label{fig-fidu-tau}Vertical optical depth of the disk at
    two wavelengths in the fiducial model. Left: Without
    bouncing. Right: With bouncing. Shown as the dashed brown
    line is the average degree of dust settling. This line uses the axis
    labeling on the right-hand side of the plot.}
\end{figure*}

Figure~\ref{fig-fidu-collision-velocities} shows the mean collision velocities
that appear in the model calculations. In both cases, the velocities are
dominated by turbulent motions, except for the outermost parts of the disk. As
expected, the mean collision velocities in the model without bouncing get close
to the fragmentation barrier. The fragmentation happens at the upper end of the
size distribution, where the relative velocities are the highest. When the
bouncing barrier is included in the model, the collision velocities are reduced
significantly because the grains do not grow as big as without the bouncing
barrier. It is noteworthy that the collision velocities exceed the bouncing
velocity (purple curve) significantly. This is because the Maxwell-Boltzmann
velocity distribution (Eq.~\ref{eq-maxwell-boltzmann}) always
has a non-zero probability for \rev{collisions} with a velocity \rev{that
is sufficiently low
for growth to be possible.} The bouncing barrier therefore never completely
inhibits growth, although the larger the grains grow, the less likely it is that a
collision leads to growth. The size distribution therefore tends to "overshoot"
the estimated bouncing barrier somewhat. To verify this effect, we show in Fig.~\ref{fig-fidu-sharpbounce}
the same model but with the Maxwell-Boltzmann velocity distribution replaced with
the mean velocity, {for both the bouncing barrier and the fragmentation
barrier}. {We found that in this case} the mean collision velocity is substantially closer
to the analytic expectation, {confirming our suspicion}. {We also
  did an experiment where we switched off the fragmentation completely just
  to be sure that a small amount of fragmentation of some grains could not
  lead to the growth of others through accretion of small dust grains. But as expected,
  because the fragmentation barrier is so much above the bouncing barrier in
  this model, the experiment did not show any difference, demonstrating that
  fragmentation does not play a role in this overshoot phenomenon.}

When dust aggregates "hit" the bouncing barrier, their growth is
inhibited, but they continue to regularly collide with each other. We
demonstrate this in Fig.~\ref{fig-fidu-nr-of-collisions} by computing
the number of collisions per megayear. With the exception of the outer edge of the
disk, the number of collisions is significant everywhere, from hundreds
to about one hundred thousand.  We therefore expected that the dust aggregates
in the bouncing barrier would be gently compressed by the
many collisions over the course of their evolution, similar to what
was demonstrated in the model of \citet{2010A&A...513A..57Z}.

The fact that the size distribution becomes so narrow in the bouncing
case, as seen in Figs.~\ref{fig-fidu-sigma} and
\ref{fig-fidu-slice-sigma}, opens up the question if the observational
appearance of disks would be strongly modified by this process.  In
order to assess this, we computed the vertical optical depth in the H
band (1.65\,\mum) and at ALMA Band 6 (1.3\,mm) wavelengths. The
results are shown in Fig.~\ref{fig-fidu-tau}. Maybe somewhat
surprisingly, the optical depth of the disk is not very strongly
affected by this change. For the ALMA wavelengths, this is caused by
the fact that the particles are not much larger than the observation
wavelength in either case, so the observations mostly probe the
mass and not the surface of the grains. Also shown in
Fig.~\ref{fig-fidu-tau} is the average degree of dust settling. In
both simulations, $H_{\mathrm{d}}/H_{\mathrm{g}}$ slowly decrease as a
function of distance from the star, with typical values of a few times
10$^{-1}$.

\begin{figure}
  \includegraphics[width=.45\textwidth]{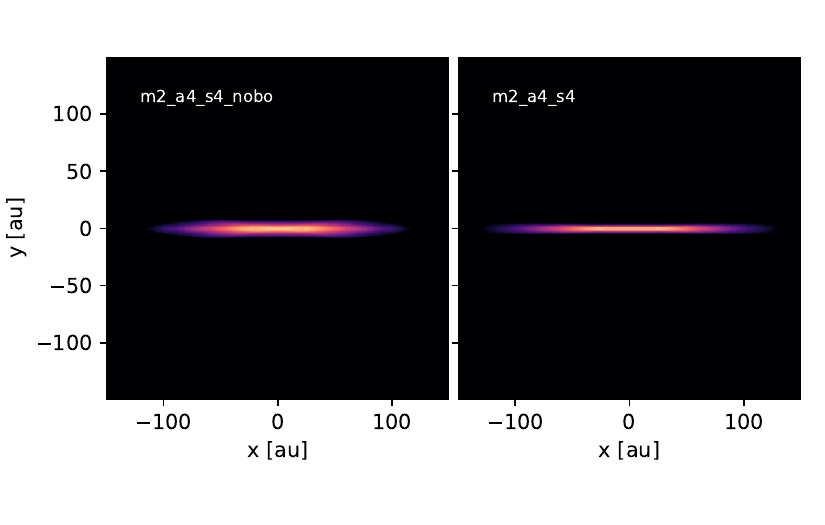}
  \caption{\label{fig-fidu-image-1.3mm}Synthetic image of the disk of the
    fiducial model seen edge-on ($i=90^\circ$) at a wavelength $\lambda=1.3$
    mm.}
\end{figure}

{We computed images at submillimeter and near-IR wavelengths using
  RADMC-3D \citep{2012ascl.soft02015D}. Dust opacities were computed
  with \texttt{optool} \citep{2021ascl.soft04010D}. For
  details on the dust opacities, see Section \ref{sec:optical-depth}.
  For the vertical distribution of the different dust grain sizes, we
  used the formula given by Equations 24 and 25 in
  \citet{2009A&A...496..597F}}. Figure~\ref{fig-fidu-image-1.3mm} shows
an edge-on ALMA view of our disk models. The model including bouncing
is slightly thinner because of the strong reduction in smaller grains
that would diffuse up vertically under some degree of turbulence.
Apart from the difference in thickness, the images are very similar.

\begin{figure}
  \includegraphics[width=.45\textwidth]{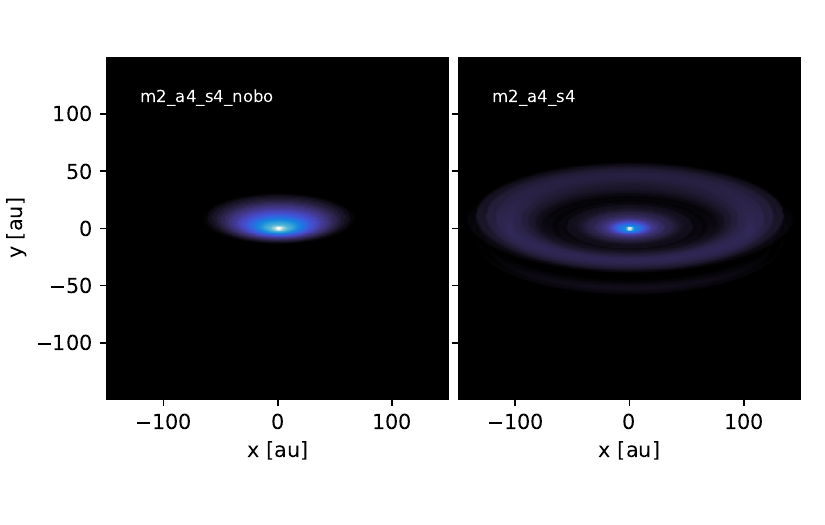}
  \caption{\label{fig-fidu-image-1-65mic}Synthetic image of the disk of the
    fiducial model seen at an inclination $i=70^\circ$ and at a wavelength
    $\lambda=1.65\mu$m.}
\end{figure}

For the shorter wavelength, the optical depth is reduced by a factor of
10-100 but not nearly enough to render the disk optically thin.
There is nevertheless a very significant effect on the appearance of
the disk in high contrast scattered light images, as shown in
Fig.~\ref{fig-fidu-image-1-65mic}. Here, bouncing leads to a strong
depression of the surface brightness in the range between 50 and 80
au. This darkening is caused by shadowing. In the inner disk, the
grains are sufficiently coupled to the gas to create a region of
higher aspect ratio. In the dark region, the absence of micron-sized
grains means that the settling of all dust is sufficient to fall into the
shadow cast by the inter region, creating what looks like a
gap. Beyond 80 au, for the case with bouncing, {the disk flaring in
combination with the small grain sizes at large distances cause
the disk surface to climb} out of the shadow again. The effect is the opposite
in the model without bouncing because small grains are present
everywhere. The inner disk becomes effectively thicker, and the disk
flares nicely to a distance of about 60 au. Farther away from the star,
the upper layers of the disk do become optically thin and fall into
the shadow created by the inner- and middle-disk regions.

The effect that the outer disk, or part of it, is shadowed by the
inner disk is often called "self-shadowing" because the disk casts a
shadow onto itself \citep{2004A&A...417..159D}. Such an effect is
commonly seen in observations of protoplanetary disks
\citep{2022A&A...658A.137G}, although it is usually considered in the
context of a disk's inner rim casting the shadow.

\begin{figure}
  \centerline{\includegraphics[width=.45\textwidth]{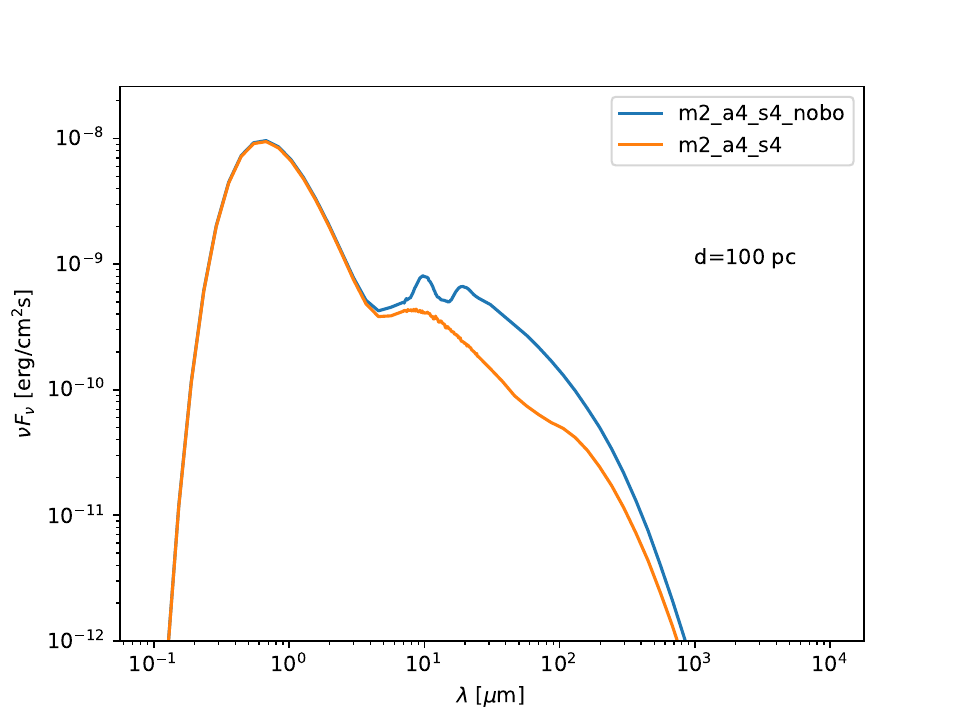}}
  \caption{\label{fig-fidu-sed}Spectral energy distribution
    of the fiducial model with and without bouncing.}
\end{figure}

{The removal of small grains also impacts the spectral energy
  distribution (SED) of the disk.  This effect is shown in
  Fig.~\ref{fig-fidu-sed}. In the model with the bouncing barrier
  turned on and fully effective, the 10 and 20\,$\mu{}$m silicate
  features that are very prominent in the no-bouncing model disappear
  entirely.  The SED at mid-IR wavelengths becomes significantly weakened,
  as the area of the disk where much of this radiation normally
  originates from becomes shadowed and colder.  It is possible that
  the effectiveness of small grain removal is too large in our
  bouncing model.  The many collisions in the bouncing regime may have
  an erosive component that would put small amounts of small grains
  back into the gas. Also, in disks with very low turbulence, the
  basic assumption of \texttt{DustPy} that full vertical mixing is
  always available and fast may be incorrect so that small grains
  are able to stay suspended higher in the disk atmosphere for a much longer
  time. Only models that treat the vertical structure of disks 
  self-consistently are able to address this point.}

In the following sections, we look at small parameter studies that
illuminate the effects of key model parameters. These parameters include the turbulent
strength $\alpha$, the monomer size $\amono$, and the disk mass
$M_{\mathrm{disk}}$.

\begin{figure*}
  \includegraphics[width=\textwidth]{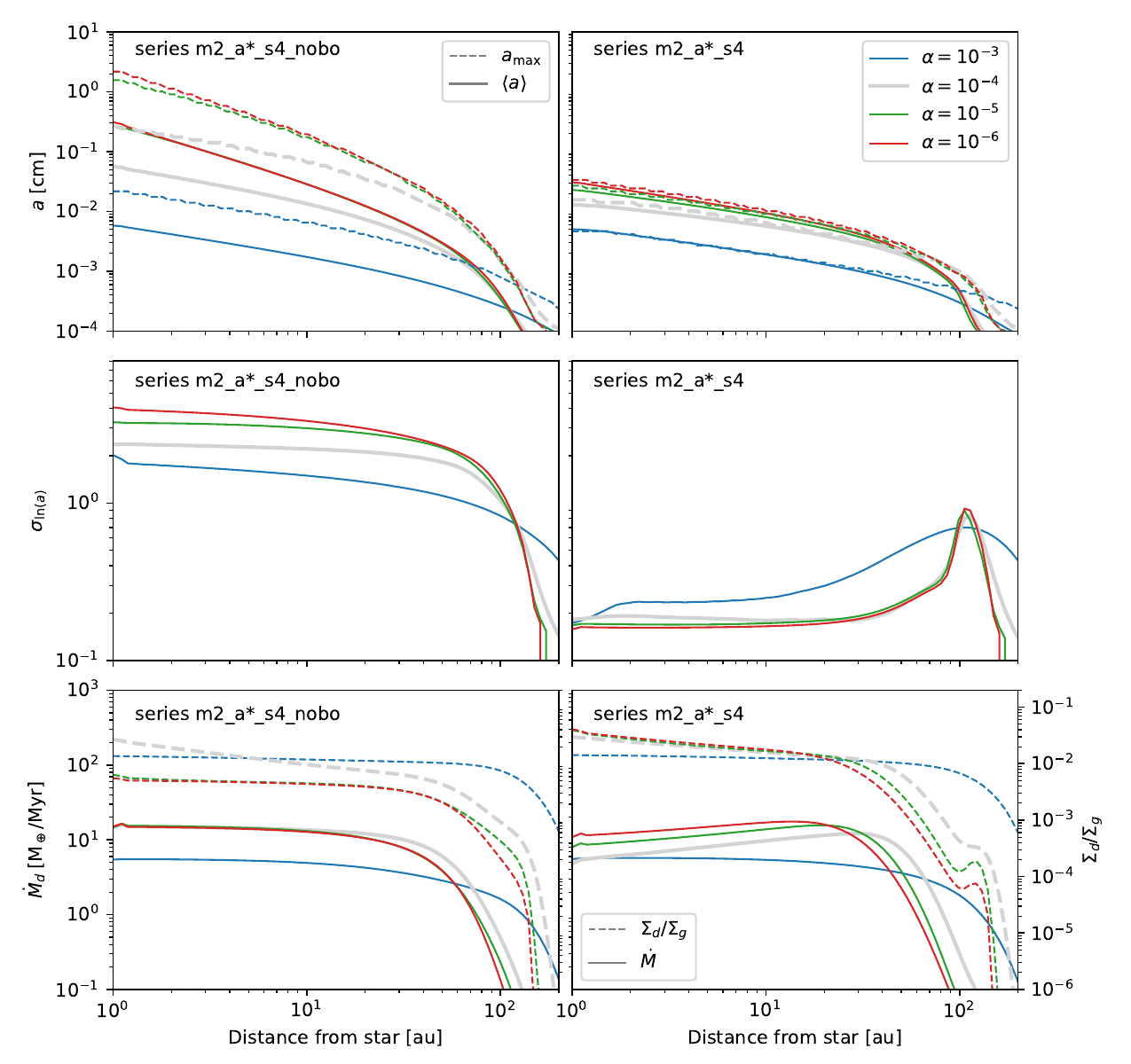}
  \caption{\label{fig-series-alpha-diagnostics}Mean particle radius,
    width of the distribution, maximum grain size {(all defined in
      section \ref{app-mean-and-width})}, and dust accretion rate
    for the \texttt{m2\_a*\_s4} series. The asterisk ($*$) stands for 3, 4, 5,
    or 6, meaning varying turbulence parameter
    $\alpha=10^{-3,4,5,6}$. The gray curves denote the fiducial model
    (\texttt{m2\_a4\_s4}).  Left: Without bouncing. Right: With
    bouncing.  }
\end{figure*}

\subsection{Model series with varying turbulent strength}\label{sec-series-alpha}

We first look at the effect of the turbulent strength on the
model result. Turbulent strength is a key parameter since there is
much uncertainty about it currently in the literature. The few
available measurements indicate values $\alpha\la2\times10^{-3}$
\citep[e.g.,][]{2015ApJ...813...99F,2016A&A...592A..49T}, but much
lower values are very seriously considered.  For this
section, we have computed models with $\alpha$ values of $10^{-3}$,
$10^{-4}$, $10^{-5}$, and $10^{-6}$. The full set of parameters for
these models can be found in Tab.~\ref{table-params}.

Turbulence is a key factor in the collisional dust growth scenario.  It is a
source of relative velocities that may, depending on strength, dominate the
relative velocities of grains. Turbulence also determines the vertical and radial
spreading of the spatial distribution of grains.

The full size distribution plots for these runs are available for
reference in Fig.~\ref{fig-series-alpha-sigmadust} in the
Appendix. Here, in Fig.~\ref{fig-series-alpha-diagnostics}, we document
the influence of changing turbulence strength on the dust size
distribution present after a megayear in a disk.  In all plots, the gray
curve represents the fiducial models discussed in Section
\ref{sec:fiducial}. The blue curve is a model with a higher turbulence,
$\alpha=10^{-3}$. The green and red curves represent the weak
turbulence cases for $\alpha=10^{-5}$ and $\alpha=10^{-6}$. 
In the figure, the upper row shows the average grain size (solid
curves) and the upper limit (dashed curves) at each location of the
disk. Without bouncing, there is a clear progression of the mean size,
increasing as the turbulence gets weaker. However, the curves for
10$^{-5}$ and 10$^{-6}$ are identical. This happens because at such
low $\alpha$, the collision velocities are determined by radial drift and no
longer by turbulence. In the case with bouncing, the mean dust size
starts to stagnate at around 100\,\mum, indicating that radial
drift already starts to take over for these smaller grains already at
$\alpha=10^{-4}$, the fiducial model.

\begin{figure*}
  \includegraphics[width=\textwidth]{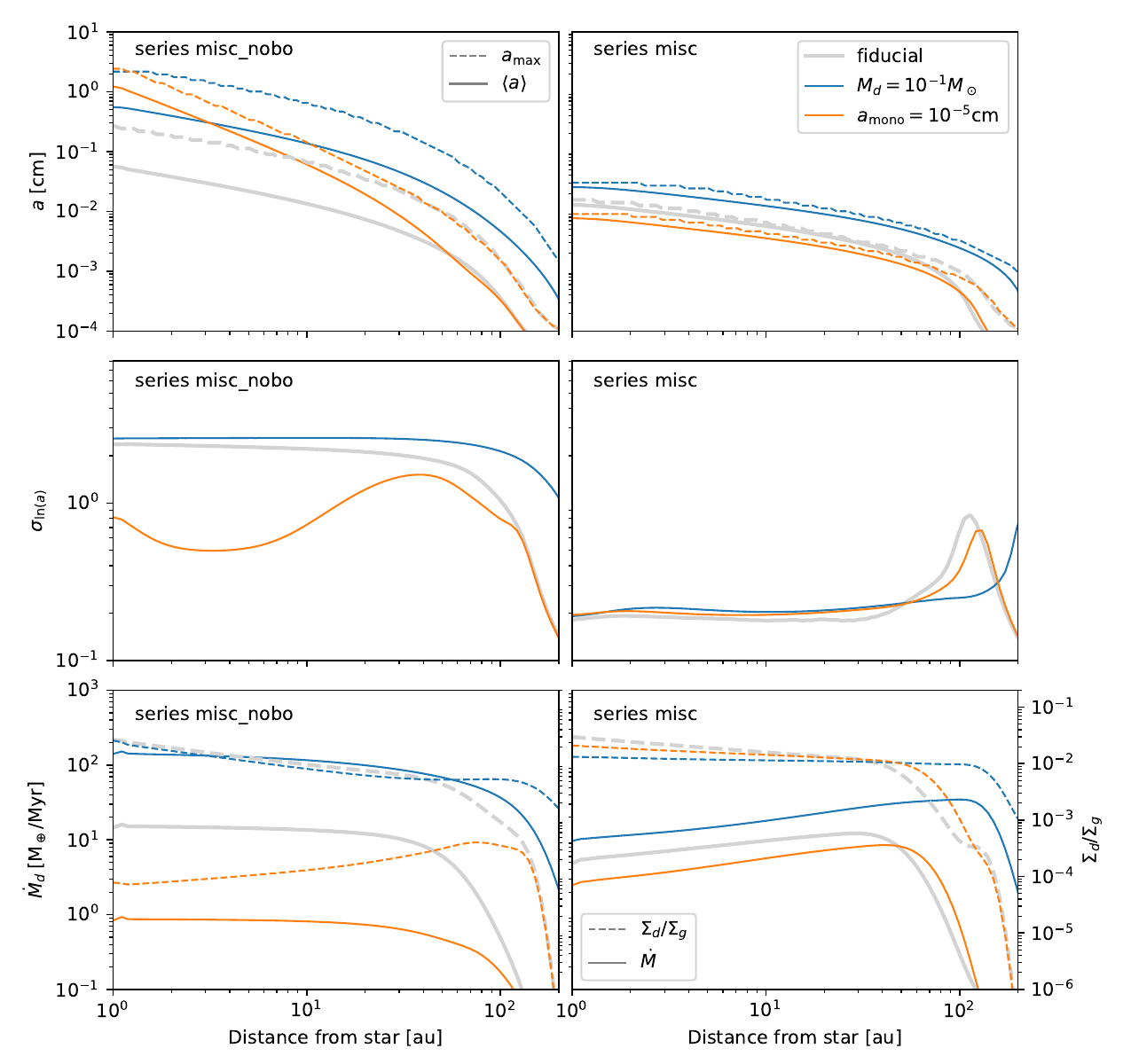}
  \caption{\label{fig-series-mixed-diagnostics}Mean particle radius, width of the
    distribution, and dust accretion rate for the
    mixed series. Left: Without bouncing. Right: With
    bouncing.
  }
\end{figure*}

The enormous differences in the width of the size distributions become
visible in the middle row of plots in
Fig.~\ref{fig-series-alpha-diagnostics}. Without bouncing, we get a
large logarithmic width of the distribution that becomes larger with
a smaller turbulence, owing to the fact that the maximum grain size
keeps increasing, while fragmentation continues to replenish very small
grains. In the case with bouncing, all but one model have a very
narrow distribution with a logarithmic width of only 0.2. Only the
largest considered value for the turbulence manages to widen the
distribution a bit because bouncing and fragmentation barriers move
closer together, and the occasional outlier in the collision velocity
produces some fragments. {This effect can be seen in
  Fig.~\ref{fig-series-alpha-sigmadust}, where the fragmentation
  barrier has moved down because of the increased relative velocities
  caused by turbulence.} The widths of the distributions peak at the
outer disk edge, where growth times are slow and the drift barrier
plays a role.

Finally, the bottom row of plots in
Fig.~\ref{fig-series-alpha-diagnostics} shows the dust mass accretion $\dot M_\mathrm{d}$,
often referred to as the "pebble flux," as a function of
distance.  Here, we can also make interesting observations.  In
the models without bouncing, the pebble flux reaches a constant,
steady state value over much of the disk. The flux is about
15 Earth masses per megayear.  In the model with the highest turbulent
speed, the value drops to 5 Earth masses per Myr because the more
violent turbulence keeps a larger fraction of the mass in smaller
grains that drift more slowly. The plot for the computation with
bouncing shows a noticeable difference.  The flux in the
$\alpha=10^{-3}$ models is quite similar to the no-bouncing case.
However, for the smaller values of alpha (leading to a very narrow
size distribution), the pebble flux is decreases toward the star
($d\dot M_\mathrm{d}/dr>0$).  What can be seen here is 
a traffic jam effect that leads to an increase
of the dust surface density and therefore an increase of the metallicity of the
disk as one gets closer to the star.

\subsection{Model series with miscellaneous varying parameters}\label{sec-series-misc}

Next, we explore two other key parameters of the models: the disk
mass $M_{\mathrm{disk}}$ and size of the monomer. A high disk mass is
often used to dial into the right conditions for making the streaming
instability and pebble accretion possible and effective. The monomer
size has an influence on the stability of aggregates and therefore on
the fragmentation velocity via Eq.~\eqref{eq:vfragscale}, so a factor
of ten reduction in monomer size leads to a very significant
increase of the fragmentation velocity by a factor of 6.8.

The full size distribution plots for these runs are available for
reference in Fig.~\ref{fig-series-mixed-sigmadust} in the
Appendix. Here, in Fig.~\ref{fig-series-mixed-diagnostics}, we show
again the fiducial model (gray line), a model where the disk mass is
increased by a factor of ten to $10^{-1}M_{\odot}$, and another model
where we decrease the monomer size. The full set of parameters for
these models can be found in Tab.~\ref{table-params}.

In the non-bouncing case, one can see that both increasing the disk
mass and reducing the monomer size leads to an increase of the mean and
maximum grain sizes. With a larger disk mass, grains are better coupled,
and collision velocities are reduced.  With a smaller monomer size,
aggregates are more rigid and harder to fragment, an effect
explaining the rise of the mean and maximum particles in this
context.  An interesting effect happens with the width of the size
distribution. The width remains large with an increased disk mass. But
in the computation with the small monomers, the size distribution suddenly
becomes very narrow, and it drops to a value of $\sim
0.45$. The explanation goes as follows:  The fragmentation velocity
increases by a factor of 6.8, so the aggregates are allowed to grow to
much larger sizes. But also the fragmentation velocity is now larger
than the drift barrier velocity so that the size distribution is now
set by the drift barrier.  In that setting, a narrow size distribution
is expected \citep{2011A&A...525A..11B}.

The changes in disk mass and monomer size have much less effect on the
outcome in the models with bouncing included. The mean and maximum
size increase by about a factor of two for the massive disk, while they
are reduced by a factor less than two in the small monomer case. The
size distributions remain narrow, indicating that in both cases,
bouncing is what sets the size distribution to almost monodisperse.

The last row of Fig.~\ref{fig-series-mixed-diagnostics} shows the
effect on the dust mass accretion rate (i.e., the pebble flux). The
changes to the pebble flux in the non-bouncing models are very
large. In the massive disk, the pebble flux is increased by almost an
order of magnitude simply because the surface density of the disk
(and consequently of the dust) is also a factor of ten higher. The model
with small monomers shows a small pebble flux after 1 Myr because
the drift has been extremely effective and the disk is already emptied
out, as most dust has already moved into the innermost disk and
evaporated. In the models with bouncing, the variations are much more
moderate, pebble fluxes change only by a factor of less than two. This
is a critically important observation.  It may mean that when the
bouncing barrier is active, increasing the disk mass is no longer the
automatic fix to create the very large pebble fluxes that are required
in many global planet formation models \citep[e.g.,][]{2019A&A...627A..83L,2019A&A...623A..88B,2021A&A...650A.152I}.

%\subsection{Dependence on monomer size}
%\subsection{Dependence on turbulence strength}
%\subsection{Dependence on dust material}
%\subsection{Production of small grains}

%\subsection{Role of the Komogorov scale}
%\subsection{Role of relative radial drift velocities}

\section{Implications and discussion}
\label{sec:implications}
We have shown that introducing the bouncing barrier into global models
of dust evolution does significantly modify the models and changes grain
sizes and grain distributions. In this section, we look at
the different areas in the context of planet formation models where
these changes may have consequences.

\subsection{Retaining dust in disks for longer}

The issue of retention of dust in disks has long been considered an
important aspect of disk models. If dust grains are allowed to grow
too quickly into the planetesimal regime and beyond, one would expect the
dust masses measured with an instrument such as ALMA to decrease on the
same timescale. If dust growth instead stops near sizes defined by Stokes numbers
close to unity, these particles are expected to drift toward the star
on a timescale of $\sim 100$ local orbital timescales. Those timescales
are too short based on the fact that we do observe massive disks out
to lifetimes of 3-20\,Myr. In addition, observations seemed
to show for a long time that the particles in disks are indeed
mittimetre to centimetre in size \citep[e.g.,][]{wilner_toward_2005}, which would have a fast
drift as an unavoidable property. Besides dust traps
\citep{pinilla_trapping_2012} or extreme porosities
\citep{kataoka_fluffy_2013}, the key mechanism to allow the
observations to be consistent with model predictions is to limit
particle growth. The various growth barriers prevent direct growth to
planetesimal sizes and keep the dust in a size range where an
instrument like ALMA is maximally sensitive
\citep{birnstiel_gas-_2010}. Also, it has become clear now that the
effects of optical depth can mimic very large grain sizes even if the
grains themselves are somewhat smaller
\citep{2014prpl.conf..339T,2022A&A...664A.137G}. This has been confirmed by
modeling polarization by self-scattering \citep{kataoka_grain_2016}. Through a
deep analysis of HD 163296 with multi-wavelength observations from
ALMA and the Karl Janski array, \citet{2022A&A...664A.137G} showed
that the grain sizes in the outer regions of this disk are on the
order of 0.2 mm.  While a
fragmentation velocity of 1\,m/s goes a long way to solving the problem
of dust retention \citep{birnstiel_dust_2009}, the bouncing barrier is
very effective in decreasing the particle sizes further and retaining
dust for longer. We observed this in
Fig.~\ref{fig-series-alpha-diagnostics}, where the 1 Myr dust surface
density is a factor of five higher in low-turbulence models with the
bouncing barrier compared to models where the barrier is
absent. Therefore, we concluded that the bouncing barrier can help
keep dust surface densities higher for longer.

\subsection{Modification of trapping efficiency}

Particles can be captured in pressure bumps that act as particle traps
\citep[e.g.,][]{2004A&A...425L...9P,2006MNRAS.373.1619R,2006A&A...453.1129P,2012A&A...538A.114P,pinilla_trapping_2012}. A
trap with any given pressure contrast can store particles in a given
size range.  Smaller particles can be extracted by turbulent mixing
and a continuing accretion flow of gas. Strong fragmentation can
make a dust trap "leaky" because the small grains produced by the
fragmentation leak out of the dust trap \citep{2023A&A...670L...5S}.

The bouncing barrier can
affect the trapping efficiency in two ways.  On the one hand, it
reduces the maximum particle size.  If the particle size set by
fragmentation is such that the particles would only be weakly trapped,
then the bouncing barrier can make the trap leaky, even for the largest
grains present.  On the other hand, the bouncing barrier leads to a
strong depletion of small micron-sized dust.  Such small particles
are normally strongly coupled to the gas and form a component that
can be pulled out of a trap by viscous mixing.  Therefore, the
bouncing barrier will make traps less leaky for small grains but
possibly more leaky for the largest grains present in the disk.

\subsection{Abundance of small grains, optical depth, and scattered
  light images}

An active bouncing barrier produces a narrow size distribution. One
consequence of this is that the amount of small grains should be
reduced to the extent that they may be completely absent. We have
shown that this is indeed what our simple bouncing barrier model
predicts.

{This may seem to contradict observations of protoplanetary disks
  at optical and mid-IR wavelengths, which do contain signatures
  of some amount of small grains. The disk does not become optically
  thin by this removal of small grains \citep{2005A&A...434..971D}
  because the upper limit of the size distribution (and also the mean)
  remain at sizes of typically 100\,\mum, rendering the disk optically
  thick at both optical and mid-IR wavelengths.}  However, we have
shown that the strong reduction in the abundance of small, \mum-sized
grains can significantly change the appearance of the disk in
scattered light by depressing parts of the disk into the shadow
produced by inner disk regions. Also, if the disk consists entirely of
100\,\mum{}-sized dust aggregates that are moderately compacted by
bouncing, then we should expect that the mid-IR spectrum does
not show a 10-\mum{} silicate feature, which is contrary to the fact
that most disks in fact show {a strong} 10-\mum{} silicate
emission feature in their spectrum
\citep{2001A&A...365..476M,2006ApJS..165..568F,2006ApJ...639..275K}.

{A perfect bouncing barrier without any production of small collisional debris
may therefore be hard to reconcile with observations of protoplanetary disks.
On the other hand, the fact that we find that each dust aggregate bounces
hundreds to thousands of times against other aggregates during the lifetime
of the disk (see Fig.~\ref{fig-fidu-nr-of-collisions}) suggests that even
a small amount of abrasion during these collisions will go some way to
replenishing the reservoir of small grains. Follow-up studies therefore should
use these observational constraints to determine how much abrasion
is necessary to keep consistent with observations and whether this will
effectively eliminate the bouncing barrier or not. A further
possibility to consider is that vertical mixing may not be as fast as
assumed in \texttt{DustPy}.  Thus, some small grains may remain in the
disk atmosphere for a long time in a more or less primordial state, as
seems to be indicated by a careful analysis of the dust in the upper
layers of the IM Lup disk
\citep{2023ApJ...944L..43T}}.

\subsection{Vertical height of the dust disk}

\begin{figure}
  \includegraphics[width=.55\textwidth]{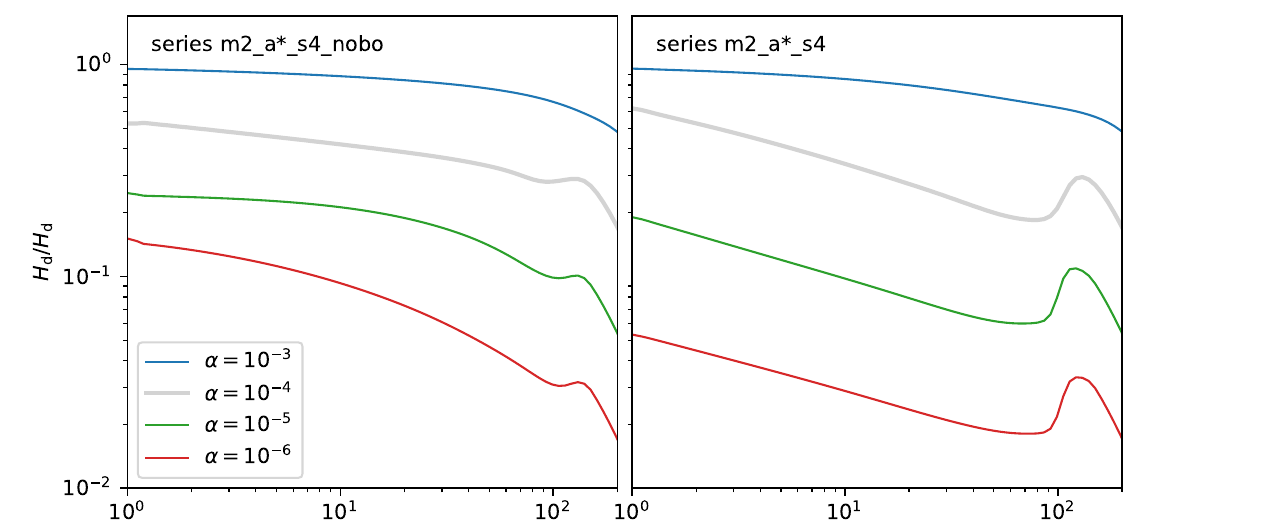}
  \caption{\label{fig-alpha-hdhg} Ratio of the dust scale height to
    the gas scale height for models with different turbulent strengths.
    Left: Without the bouncing barrier. Right: With the bouncing barrier.}
\end{figure}

Observations of edge-on planet-forming disks with ALMA have revealed
that the midplane of the disk, supposedly composed of large grains in
the millimetre to centimetre range, is very geometrically
thin. \citet{2016ApJ...816...25P} constrained the aspect ratio of the
rings in the HL Tau disk to about $H_{\mathrm{d}}/r\simeq0.01$.
\citet{2022ApJ...930...11V} measured the ratio of the dust and gas
scale heights in the disk of Oph\,163131 to be
$H_{\mathrm{d}}/H_{\mathrm{g}}\simeq
5\times10^{-2}$. \citet{1995Icar..114..237D} and
\citet{birnstiel_gas-_2010} showed that for Stokes numbers less than
one, the vertical width $H_{\mathrm{d}}$ will be related to the ratio
of the turbulent strength and the Stokes number by
\begin{equation}
\label{eq:1}
\frac{H_{\mathrm{d}}}{H_{\mathrm{g}}} =
\min\left(1,\sqrt{\frac{\alpha}{\mathrm{St}}}\right) \quad .
\end{equation}
Since the bouncing barrier reduces the upper limit of the size
distribution, and therefore the Stokes number of those grains, a model
with the bouncing barrier will place even more stringent constraints on
the turbulent mixing available.  For our modeling grid with varying
turbulent strength, we show the ratio $H_{\mathrm{d}}/H_{\mathrm{g}}$
in Fig.~\ref{fig-alpha-hdhg}. To reach the measured thickness
of the Oph\,163131 disk, we would require $\alpha\la10^{-5}$.  We note
that his specific result still depends on the adopted gas surface
density.

\subsection{Streaming instability and planetesimal formation, pebble
  accretion}
\label{sec:si-pf-pa}
The streaming instability is currently the favorite model to explain
the formation of planetesimals. In order to activate the streaming
instability, a number of criteria need to be fulfilled. Roughly
speaking, these criteria are (1) the local metallicity of the disk,
which must be (depending on the study) not too far below solar
\citep{li_thresholds_2021} or significantly above solar
\citep{johansen_particle_2009}, and (2) the Stokes number of the
pebbles drifting through the gas. Stokes numbers between 10$^{-3}$ and
1 have been reported to be effective. {These criteria apply strictly to
  laminar disks. For a disk that has active turbulence, the efficiency of
  the streaming instability can be reduced or even become zero if
  grains are not allowed to settle to the midplane
  \citep{2020ApJ...904..132G}.} Recently, there has been a lot of discussion
  on the effects of a polydisperse \citep{krapp_streaming_2019} or
  even continuous \citep{paardekooper_polydisperse_2020} size
  distributions because of the expectation derived from models with
  the fragmentation barrier that size distributions in disks are
  broad.

\begin{figure*}[t]
  \begin{center}
  \includegraphics[width=.48\textwidth]{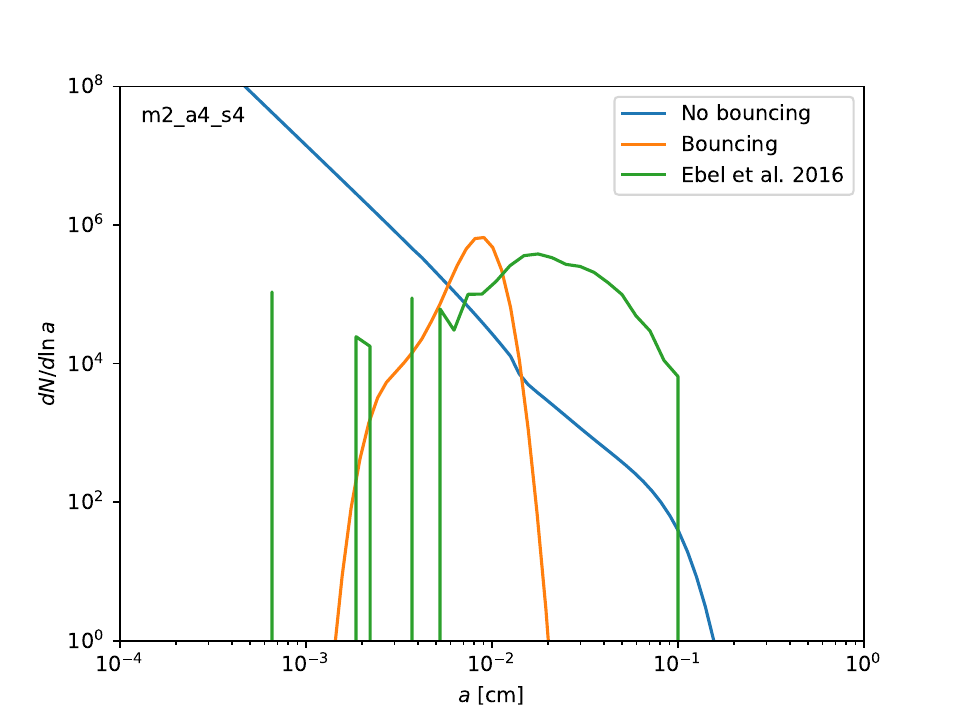}
  \includegraphics[width=.48\textwidth]{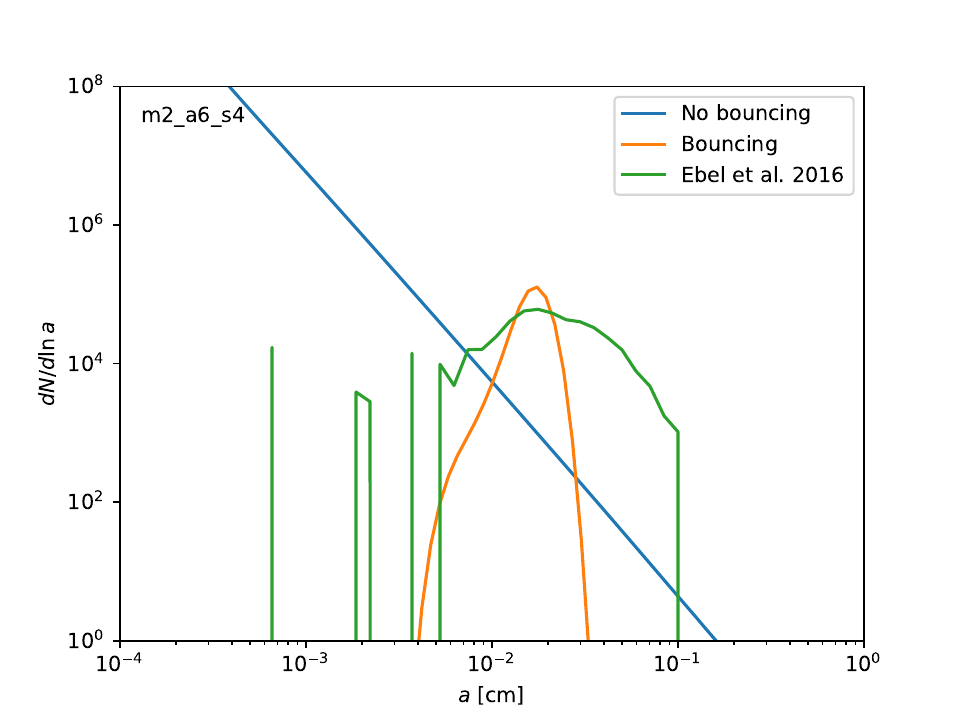}
  \end{center}
  \caption{\label{fig-compare-chondrules}Logarithmic size distribution
    ($a\,dN/da=dN/\ln a$) of dust aggregates at $r=2.9\;\mathrm{au}$ in the
    protoplanetary disk. Left: Fiducial model \texttt{m2\_a4\_s4}. Right: Model
    \texttt{m2\_a6\_s4}, which is like the fiducial model but with a 100-times lower
    $\alpha$ value. The blue and orange curves are for the cases without a bouncing
    barrier and with a bouncing barrier, respectively, and they have been normalized
    such that $N=\int_0^\infty dN$ is the number of aggregates per
    square centimetre of the disk. The Ebel results are the chondrule radius
    distribution inferred from thin slices of Allende, shown with the green
    curve of their figure 7 (we note that we plot against the radius, while
    \citet{2016GeCoA.172..322E} plotted against the 2-log of the diameter). The
    normalization of the Ebel curve has been adjusted so that it equals that of the
    two other curves. We also note that due to the limited sample size, the
    renormalized Ebel curve is noise limited at about $dN/d\ln a\simeq 10^4$.}
\end{figure*}

The bouncing barrier changes this situation in the following ways.
First of all, we have shown that with an active bouncing barrier, a
very narrow size distribution is obtained. A monodisperse size
distribution is a reasonable approximation of this, 
and the modifications required for polydisperse or continuous size
distributions may not be necessary. Second, the bouncing barrier reduces
the typical size of the pebbles present in the disk, leading to a
smaller Stokes number. That does not exclude the activation of the
streaming instability, but it will slow it down by extending the
growth times of the instability. \cite{li_thresholds_2021} presented an extensive parameter
study and found that the pre-clumping times typically grow by a factor
of 100 when the Stokes number of the particle drops from 1 to
10$^{-2}$ or 10$^{-3}$. In some cases, for the lower Stokes numbers,
no clumping is observed.  Therefore, we concluded that the bouncing
barrier will significantly affect the streaming instability by
providing an essentially monodisperse size distribution and reduced
Stokes numbers.

As for the efficiency of the pebble accretion scenario, we have shown in this
paper that the bouncing barrier appears to prevent the formation of the very
high pebble flux $\dot M_{\mathrm{d}}$ needed for the standard models of pebble
accretion \citep[see e.g.,][]{2019A&A...623A..88B}. The presence of the bouncing
barrier thus appears to be detrimental to planet formation.  However, pebble
accretion can also occur in dust traps or dust rings, where the efficiency does
not depend on the pebble flux but on the pebble surface density
\citep[e.g.,][]{2023MNRAS.518.3877J}. The effect of the bouncing barrier
on the efficiency of planet formation may thus not be as simple as comparing
the magnitude of the pebble flux.

\subsection{Chondrule formation}\label{sec-chondrule-formation}
The narrow size distribution of dust aggregates resulting from the
bouncing barrier may have an interesting implication for chondritic
meteorites.  Chondrules are once-molten "granules" (tiny spherical
rocks) that make up a substantial fraction of the mass of chondritic
meteorites. They have typical diameters between 0.1 and 1 millimeter,
dependent on their chondrite group \citep{2012M&PS...47.1176J}. A summary of
measured size distributions of chondrules can be found in
\citet{2015ChEG...75..419F}. The formation mechanism of chondrules is
still a matter of debate. The mystery lies in the fact that they are
about 100$\cdots$1000 times larger and $10^{6\cdots 9}$ times more
massive than interstellar dust grains, yet they are solid rocks
instead of loosely bound dust aggregates. From petrological studies it
can be inferred that during the first few million years of the Solar
System, they must have undergone one or more "flash heating" events
\citep{1996GeCoA..60.3115J}, causing a dust aggregate to melt and form
a tiny spherical rock that we today identify as a chondrule.

One possible chondrule formation mechanism is the passing of dust aggregates through the
bow-shocks of planetary embryos on eccentric orbits
\citep{2012ApJ...752...27M}. In this scenario, dust aggregates that
freely float in the protoplanetary disk may have chance encounters
with a planetary embryo that moves supersonically through the disk due
to high eccentricity. The probability that the dust aggregate accretes
onto the embryo is very low due to the small cross section of the
embryo. But the probability of passing through the bow-shock induced
by the embryo in the protoplanetary disk is much higher. In that case,
the dust aggregate gets heated to temperatures above the liquidus and
subsequently cools back to the background temperature to form a chondrule.

However, \rev{a key question} is why the chondrules in meteorites are limited to a narrow size
range between 0.1 and 1 millimeter. In terms of this scenario, this
question translates into asking why the precursor dust
  aggregates have such a narrow size distribution. Turbulent
size sorting has been cited and extensively modeled as a possible
explanation \citep[e.g.,][]{2001ApJ...546..496C}. {That scenario,
  however, does not explain what happens to the de-selected dust aggregates.}
{An alternative explanation was given 
  by \citet{2014Icar..232..176J}. He noted that the bouncing barrier has a
  very weak dependence on disk parameters, such as the turbulence
  parameter $\alpha$, leading to a rather model-independent maximum
  dust aggregate size and a narrow size distribution.
  If these dust aggregates are then flash heated,
  they  produce the right kind of distributions of sizes for the resulting
  chondrules.}

{We found exactly this phenomenon happening in our models. As can
  be seen in Fig.~\ref{fig-series-alpha-sigmadust}, even for strongly
  varying $\alpha$, the typical sizes of the dust aggregates remain
  rather similar. Indeed, we already found this insensitivity in our
  analytic estimates of the bouncing barrier in Section \ref{sec:estimate-bb},
  cf.~Eq.~(\ref{eq:abouncezero}).} As we show in this paper, and as can also be seen in
\cite{2016ApJ...818..200E}, the presence of the bouncing barrier
produces a very narrow size distribution of dust aggregates. If only a
certain fraction of these dust aggregates are flash heated, the
remainder of the dust aggregates stay unmolten and are available to
form the matrix of a chondrite in a planetesimal-forming event, such
as those triggered by the streaming instability
\citep{2015A&A...579A..43C}.

To test this hypothesis, we compared the vertically integrated size
distribution of our fiducial model (\texttt{m2\_a4\_s4} and its
no-bouncing sibling) at $r=2.9\;\mathrm{au}$ with the chondrule size
distribution in the Allende family of chondritic meteorites. We used
the size distribution determined by \citet{2016GeCoA.172..322E}, as
shown in the green curve of their Figure 7. They do not count matrix
particles, but their results show a clear scarcity of particles with
radii between $10^{-3}$ and $10^{-2}$ cm and a clear peak of the
distribution at around a radius of 0.15 mm. The results of the
comparison are shown in Fig.~\ref{fig-compare-chondrules} (left).

It is evident that the model without the bouncing barrier (blue
curve in Fig.~\ref{fig-compare-chondrules}) has a problem when compared to the chondrule statistics (the
green curve).  It predicts far too many small particles with sizes
between monomers and 0.1 mm, and it does not have a peak at some
typical size. However, it does have a cutoff radius very similar to
that of the chondrules. The model with the bouncing barrier (the
orange curve) also has problems. It is narrower than the
chondrule statistics of Ebel et al., and it peaks at a too small radius
(by about a factor of two).

In Fig.~\ref{fig-compare-chondrules} (right) the same is shown but for a much
lower turbulent strength: $\alpha=10^{-6}$. The model without the bouncing
barrier in figure differs even more from the chondrule curve, and the model with the
bouncing barrier peaks at the correct radius, yet it is still too narrow.

This comparison has to be interpreted with some care. The radii of the
dust aggregate precursors of the chondrules are porous. In our model,
we have a porosity of about 0.397, implying an increased radius for
the same mass of about 15\%, which is still small compared to the
range of radii over which we compare the results. The porosity of the
actual chondrule precursors may have been higher. Furthermore, other
parameters, such as the disk mass and the monomer size, enter into the
result as well. But overall, the rough size range seems to match, even
though the observed size distribution is not quite as narrow as our
calculations predict. The bouncing barrier therefore should be
considered as a possible factor contributing to the size distribution
of chondrules in meteorites.

\section{Conclusions}
\label{sec:conclusions}

We studied the effects of the bouncing barrier in global models
of dust evolution in planet-forming disks.  Through our work, we arrived at the following
conclusions:

\begin{enumerate}
\item\label{item:1} When the bouncing is implemented into a global
  dust evolution model, it introduces several important changes.
\item The dust size distribution becomes much narrower, with a
  standard deviation in the natural logarithm of only 0.2, compared to
  a formal width of 2.5 in a typical non-bouncing run. The dust is then
  present as a quasi-monodisperse size set by the bouncing barrier.
\item The efficient removal of small \mum-sized grains should be
  visible in scattered light images and mid-IR spectroscopy,
  and it might be diagnosed by them.
\item Nevertheless, the disk does not become vertically optically thin
  unless much dust is converted into planetesimals.
\item Both the mean and the maximum steady-state dust particle sizes
  change significantly when the bouncing barrier is activated. The
  amount of change depends on a number of parameters, such as the
  turbulence strength and the disk mass.
\item The bouncing barrier strongly dampens the effect of changing the
  disk mass. Without bouncing, an increase in disk mass leads to a
  larger pebble flux initially and then to a more rapid global dust
  depletion. Bouncing moderates both of these effects. The pebble flux
  increases only moderately, and the dust disk survives longer.
\item The effects of making the dust aggregates mechanically (for
  example by using a smaller monomer size, which in effect increases the
  fragmentation velocity) are also very much damped with bouncing.
\item In the presence of an active bouncing barrier, the streaming
  instability will likely not be polydisperse because of the quasi-monodisperse 
  size distribution. The speed and the efficiency of
  the streaming instability will be modified, but details will depend
  on circumstances.
\item The narrow size distribution of chondrules found in individual
  meteorites may be related to the bouncing barrier because the
  bouncing barrier may set the population of dust aggregates that are
  chondrule precursors.
\item {The universality of chondrule sizes across different meteorites could
  be caused by the insensitivity of the bouncing barrier to disk parameters,
  in line with what was suggested by \citet{2014Icar..232..176J}.}
\end{enumerate}

The bouncing barrier, when effective, does change the course of the
evolution of the dust component in a planet-forming disk. In particular, the barrier should be
considered in models that heavily lean on streaming
instability, particle concentration, and pebble accretion.

\begin{acknowledgements}
  This paper has been made possible by the excellent \texttt{DustPy}
  modelling tool, written and made available by Sebastian Stammler and
  Til Birnstiel at \url{https://github.com/stammler/dustpy}. Kudos and
  thanks to the authors!  We are grateful to the anonymous referee who
  provided a very timely and detailed report, helping to clarify a
  number of key aspects. We acknowledge funding from the DFG research
  group FOR 2634 ``Planet Formation Witnesses and Probes: Transition
  Disks'' under grant DU 414/28-1, and from the DFG priority programm
  SPP 1992 “Diversity of Extrasolar Planets” (funding the summer
  school on planet formation in August 2023, where part of this work
  was written).
\end{acknowledgements}

%-------------------------------------------------------------------

\bibliographystyle{aa}
\bibliography{ms}

\begin{thebibliography}{102}
\expandafter\ifx\csname natexlab\endcsname\relax\def\natexlab#1{#1}\fi

\bibitem[{Birnstiel {et~al.}(2009)Birnstiel, Dullemond, \&
  Brauer}]{birnstiel_dust_2009}
Birnstiel, T., Dullemond, C.~P., \& Brauer, F. 2009, Astronomy and
  Astrophysics, 503, L5, aDS Bibcode: 2009A\&A...503L...5B

\bibitem[{Birnstiel {et~al.}(2010)Birnstiel, Dullemond, \&
  Brauer}]{birnstiel_gas-_2010}
Birnstiel, T., Dullemond, C.~P., \& Brauer, F. 2010, Astronomy and
  Astrophysics, Volume 513, id.A79,
  {\textless}NUMPAGES{\textgreater}21{\textless}/NUMPAGES{\textgreater} pp.,
  513, A79

\bibitem[{{Birnstiel} {et~al.}(2012){Birnstiel}, {Klahr}, \&
  {Ercolano}}]{2012A&A...539A.148B}
{Birnstiel}, T., {Klahr}, H., \& {Ercolano}, B. 2012, \aap, 539, A148

\bibitem[{Birnstiel {et~al.}(2011)Birnstiel, Ormel, \&
  Dullemond}]{2011A&A...525A..11B}
Birnstiel, T., Ormel, C.~W., \& Dullemond, C.~P. 2011, Astronomy and
  Astrophysics, 525, 11

\bibitem[{{Bitsch} {et~al.}(2019){Bitsch}, {Izidoro}, {Johansen}, {Raymond},
  {Morbidelli}, {Lambrechts}, \& {Jacobson}}]{2019A&A...623A..88B}
{Bitsch}, B., {Izidoro}, A., {Johansen}, A., {et~al.} 2019, \aap, 623, A88

\bibitem[{Booth {et~al.}(2018)Booth, Meru, Lee, \&
  Clarke}]{booth_breakthrough_2018}
Booth, R.~A., Meru, F., Lee, M.~H., \& Clarke, C.~J. 2018, Monthly Notices of
  the Royal Astronomical Society, 475, 167, aDS Bibcode: 2018MNRAS.475..167B

\bibitem[{{Brauer} {et~al.}(2008){Brauer}, {Dullemond}, \&
  {Henning}}]{2008A&A...480..859B}
{Brauer}, F., {Dullemond}, C.~P., \& {Henning}, T. 2008, \aap, 480, 859

\bibitem[{Brisset {et~al.}(2017)Brisset, Heißelmann, Kothe, Weidling, \&
  Blum}]{brisset_low-velocity_2017}
Brisset, J., Heißelmann, D., Kothe, S., Weidling, R., \& Blum, J. 2017,
  Astronomy and Astrophysics, 603, A66

\bibitem[{{Carrera} {et~al.}(2015){Carrera}, {Johansen}, \&
  {Davies}}]{2015A&A...579A..43C}
{Carrera}, D., {Johansen}, A., \& {Davies}, M.~B. 2015, \aap, 579, A43

\bibitem[{Chokshi {et~al.}(1993)Chokshi, Tielens, \&
  {Hollenbach}}]{chokshi_dust_1993}
Chokshi, A., Tielens, A., \& {Hollenbach}. 1993, Astrophysical Journal, 407,
  806

\bibitem[{{Cuzzi} {et~al.}(2001){Cuzzi}, {Hogan}, {Paque}, \&
  {Dobrovolskis}}]{2001ApJ...546..496C}
{Cuzzi}, J.~N., {Hogan}, R.~C., {Paque}, J.~M., \& {Dobrovolskis}, A.~R. 2001,
  \apj, 546, 496

\bibitem[{{Dominik} {et~al.}(2021){Dominik}, {Min}, \&
  {Tazaki}}]{2021ascl.soft04010D}
{Dominik}, C., {Min}, M., \& {Tazaki}, R. 2021, {OpTool: Command-line driven
  tool for creating complex dust opacities}, Astrophysics Source Code Library,
  record ascl:2104.010

\bibitem[{Dominik \& Tielens(1997)}]{dominik_physics_1997}
Dominik, C. \& Tielens, A. G. G.~M. 1997, The Astrophysical Journal, 480, 647,
  aDS Bibcode: 1997ApJ...480..647D

\bibitem[{{Dominik} \& {Tielens}(1997)}]{1997ApJ...480..647D}
{Dominik}, C. \& {Tielens}, A.~G.~G.~M. 1997, \apj, 480, 647

\bibitem[{{Dorschner} {et~al.}(1995){Dorschner}, {Begemann}, {Henning},
  {Jaeger}, \& {Mutschke}}]{1995A&A...300..503D}
{Dorschner}, J., {Begemann}, B., {Henning}, T., {Jaeger}, C., \& {Mutschke}, H.
  1995, \aap, 300, 503

\bibitem[{{Drazkowska} {et~al.}(2022){Drazkowska}, {Bitsch}, {Lambrechts},
  {Mulders}, {Harsono}, {Vazan}, {Liu}, {Ormel}, {Kretke}, \&
  {Morbidelli}}]{Drazkowska_2022_ppvii}
{Drazkowska}, J., {Bitsch}, B., {Lambrechts}, M., {et~al.} 2022, arXiv
  e-prints, arXiv:2203.09759

\bibitem[{Drążkowska {et~al.}(2013)Drążkowska, Windmark, \&
  Dullemond}]{drazkowska_planetesimal_2013}
Drążkowska, J., Windmark, F., \& Dullemond, C.~P. 2013, Astronomy and
  Astrophysics, 556, A37

\bibitem[{{Dubrulle} {et~al.}(1995){Dubrulle}, {Morfill}, \&
  {Sterzik}}]{1995Icar..114..237D}
{Dubrulle}, B., {Morfill}, G., \& {Sterzik}, M. 1995, \icarus, 114, 237

\bibitem[{{Dullemond} \& {Dominik}(2004)}]{2004A&A...417..159D}
{Dullemond}, C.~P. \& {Dominik}, C. 2004, \aap, 417, 159

\bibitem[{{Dullemond} \& {Dominik}(2005)}]{2005A&A...434..971D}
{Dullemond}, C.~P. \& {Dominik}, C. 2005, \aap, 434, 971

\bibitem[{{Dullemond} {et~al.}(2012){Dullemond}, {Juhasz}, {Pohl}, {Sereshti},
  {Shetty}, {Peters}, {Commercon}, \& {Flock}}]{2012ascl.soft02015D}
{Dullemond}, C.~P., {Juhasz}, A., {Pohl}, A., {et~al.} 2012, {RADMC-3D: A
  multi-purpose radiative transfer tool}, Astrophysics Source Code Library,
  record ascl:1202.015

\bibitem[{{Ebel} {et~al.}(2016){Ebel}, {Brunner}, {Konrad}, {Leftwich}, {Erb},
  {Lu}, {Rodriguez}, {Crapster-Pregont}, {Friedrich}, \&
  {Weisberg}}]{2016GeCoA.172..322E}
{Ebel}, D.~S., {Brunner}, C., {Konrad}, K., {et~al.} 2016, \gca, 172, 322

\bibitem[{{Estrada} \& {Cuzzi}(2022)}]{2022ApJ...936...40E}
{Estrada}, P.~R. \& {Cuzzi}, J.~N. 2022, \apj, 936, 40

\bibitem[{{Estrada} {et~al.}(2016){Estrada}, {Cuzzi}, \&
  {Morgan}}]{2016ApJ...818..200E}
{Estrada}, P.~R., {Cuzzi}, J.~N., \& {Morgan}, D.~A. 2016, \apj, 818, 200

\bibitem[{{Estrada} {et~al.}(2022){Estrada}, {Cuzzi}, \&
  {Umurhan}}]{2022ApJ...936...42E}
{Estrada}, P.~R., {Cuzzi}, J.~N., \& {Umurhan}, O.~M. 2022, \apj, 936, 42

\bibitem[{{Flaherty} {et~al.}(2015){Flaherty}, {Hughes}, {Rosenfeld},
  {Andrews}, {Chiang}, {Simon}, {Kerzner}, \& {Wilner}}]{2015ApJ...813...99F}
{Flaherty}, K.~M., {Hughes}, A.~M., {Rosenfeld}, K.~A., {et~al.} 2015, \apj,
  813, 99

\bibitem[{{Friedrich} {et~al.}(2015){Friedrich}, {Weisberg}, {Ebel}, {Biltz},
  {Corbett}, {Iotzov}, {Khan}, \& {Wolman}}]{2015ChEG...75..419F}
{Friedrich}, J.~M., {Weisberg}, M.~K., {Ebel}, D.~S., {et~al.} 2015, Chemie der
  Erde / Geochemistry, 75, 419

\bibitem[{{Fromang} \& {Nelson}(2009)}]{2009A&A...496..597F}
{Fromang}, S. \& {Nelson}, R.~P. 2009, \aap, 496, 597

\bibitem[{{Furlan} {et~al.}(2006){Furlan}, {Hartmann}, {Calvet}, {D'Alessio},
  {Franco-Hern{\'a}ndez}, {Forrest}, {Watson}, {Uchida}, {Sargent}, {Green},
  {Keller}, \& {Herter}}]{2006ApJS..165..568F}
{Furlan}, E., {Hartmann}, L., {Calvet}, N., {et~al.} 2006, \apjs, 165, 568

\bibitem[{{Garufi} {et~al.}(2022){Garufi}, {Dominik}, {Ginski}, {Benisty}, {van
  Holstein}, {Henning}, {Pawellek}, {Pinte}, {Avenhaus}, {Facchini},
  {Galicher}, {Gratton}, {M{\'e}nard}, {Muro-Arena}, {Milli}, {Stolker},
  {Vigan}, {Villenave}, {Moulin}, {Origne}, {Rigal}, {Sauvage}, \&
  {Weber}}]{2022A&A...658A.137G}
{Garufi}, A., {Dominik}, C., {Ginski}, C., {et~al.} 2022, \aap, 658, A137

\bibitem[{{Gole} {et~al.}(2020){Gole}, {Simon}, {Li}, {Youdin}, \&
  {Armitage}}]{2020ApJ...904..132G}
{Gole}, D.~A., {Simon}, J.~B., {Li}, R., {Youdin}, A.~N., \& {Armitage}, P.~J.
  2020, \apj, 904, 132

\bibitem[{{Gong} {et~al.}(2021){Gong}, {Ivlev}, {Akimkin}, \&
  {Caselli}}]{2021ApJ...917...82G}
{Gong}, M., {Ivlev}, A.~V., {Akimkin}, V., \& {Caselli}, P. 2021, \apj, 917, 82

\bibitem[{{Guidi} {et~al.}(2022){Guidi}, {Isella}, {Testi}, {Chandler}, {Liu},
  {Schmid}, {Rosotti}, {Meng}, {Jennings}, {Williams}, {Carpenter}, {de
  Gregorio-Monsalvo}, {Li}, {Liu}, {Ortolani}, {Quanz}, {Ricci}, \&
  {Tazzari}}]{2022A&A...664A.137G}
{Guidi}, G., {Isella}, A., {Testi}, L., {et~al.} 2022, \aap, 664, A137

\bibitem[{Gundlach \& Blum(2015)}]{gundlach_stickiness_2015}
Gundlach, B. \& Blum, J. 2015, The Astrophysical Journal, 798, 34, aDS Bibcode:
  2015ApJ...798...34G

\bibitem[{Güttler {et~al.}(2010)Güttler, Blum, Zsom, Ormel, \&
  Dullemond}]{guttler_outcome_2010}
Güttler, C., Blum, J., Zsom, A., Ormel, C.~W., \& Dullemond, C.~P. 2010,
  Astronomy and Astrophysics, 513, A56

\bibitem[{{Hartlep} \& {Cuzzi}(2020)}]{2020ApJ...892..120H}
{Hartlep}, T. \& {Cuzzi}, J.~N. 2020, \apj, 892, 120

\bibitem[{{Heim} {et~al.}(1999){Heim}, {Blum}, {Preuss}, \&
  {Butt}}]{1999PhRvL..83.3328H}
{Heim}, L.-O., {Blum}, J., {Preuss}, M., \& {Butt}, H.-J. 1999, \prl, 83, 3328

\bibitem[{Hill {et~al.}(2015)Hill, Heißelmann, Blum, \&
  Fraser}]{hill_collisions_2015}
Hill, C.~R., Heißelmann, D., Blum, J., \& Fraser, H.~J. 2015, Astronomy and
  Astrophysics, 573, A49

\bibitem[{Izidoro {et~al.}(2021)Izidoro, Bitsch, Raymond, Johansen, Morbidelli,
  Lambrechts, \& Jacobson}]{izidoro_formation_2021}
Izidoro, A., Bitsch, B., Raymond, S.~N., {et~al.} 2021, Astronomy and
  Astrophysics, 650, A152, aDS Bibcode: 2021A\&A...650A.152I

\bibitem[{{Izidoro} {et~al.}(2021){Izidoro}, {Bitsch}, {Raymond}, {Johansen},
  {Morbidelli}, {Lambrechts}, \& {Jacobson}}]{2021A&A...650A.152I}
{Izidoro}, A., {Bitsch}, B., {Raymond}, S.~N., {et~al.} 2021, \aap, 650, A152

\bibitem[{{Jacquet}(2014)}]{2014Icar..232..176J}
{Jacquet}, E. 2014, \icarus, 232, 176

\bibitem[{{Jiang} \& {Ormel}(2023)}]{2023MNRAS.518.3877J}
{Jiang}, H. \& {Ormel}, C.~W. 2023, \mnras, 518, 3877

\bibitem[{{Johansen} {et~al.}(2007){Johansen}, {Oishi}, {Mac Low}, {Klahr},
  {Henning}, \& {Youdin}}]{2007Natur.448.1022J}
{Johansen}, A., {Oishi}, J.~S., {Mac Low}, M.-M., {et~al.} 2007, \nat, 448,
  1022

\bibitem[{{Johansen} \& {Youdin}(2007)}]{2007ApJ...662..627J}
{Johansen}, A. \& {Youdin}, A. 2007, \apj, 662, 627

\bibitem[{Johansen {et~al.}(2009)Johansen, Youdin, \&
  Mac~Low}]{johansen_particle_2009}
Johansen, A., Youdin, A., \& Mac~Low, M.-M. 2009, The Astrophysical Journal,
  704, L75, aDS Bibcode: 2009ApJ...704L..75J

\bibitem[{{Jones}(1996)}]{1996GeCoA..60.3115J}
{Jones}, R.~H. 1996, \gca, 60, 3115,3120

\bibitem[{{Jones}(2012)}]{2012M&PS...47.1176J}
{Jones}, R.~H. 2012, \maps, 47, 1176

\bibitem[{Kataoka {et~al.}(2016)Kataoka, Muto, Momose, Tsukagoshi, \&
  Dullemond}]{kataoka_grain_2016}
Kataoka, A., Muto, T., Momose, M., Tsukagoshi, T., \& Dullemond, C.~P. 2016,
  The Astrophysical Journal, 820, 54, aDS Bibcode: 2016ApJ...820...54K

\bibitem[{Kataoka {et~al.}(2013)Kataoka, Tanaka, Okuzumi, \&
  Wada}]{kataoka_fluffy_2013}
Kataoka, A., Tanaka, H., Okuzumi, S., \& Wada, K. 2013, Astronomy and
  Astrophysics, 557, L4

\bibitem[{Kelling {et~al.}(2014)Kelling, Wurm, \&
  Köster}]{kelling_experimental_2014}
Kelling, T., Wurm, G., \& Köster, M. 2014, The Astrophysical Journal, 783,
  111, aDS Bibcode: 2014ApJ...783..111K

\bibitem[{{Kessler-Silacci} {et~al.}(2006){Kessler-Silacci}, {Augereau},
  {Dullemond}, {Geers}, {Lahuis}, {Evans}, {van Dishoeck}, {Blake}, {Boogert},
  {Brown}, {J{\o}rgensen}, {Knez}, \& {Pontoppidan}}]{2006ApJ...639..275K}
{Kessler-Silacci}, J., {Augereau}, J.-C., {Dullemond}, C.~P., {et~al.} 2006,
  \apj, 639, 275

\bibitem[{Krapp {et~al.}(2019)Krapp, Benítez-Llambay, Gressel, \&
  Pessah}]{krapp_streaming_2019}
Krapp, L., Benítez-Llambay, P., Gressel, O., \& Pessah, M.~E. 2019, The
  Astrophysical Journal, 878, L30, aDS Bibcode: 2019ApJ...878L..30K

\bibitem[{{Krause} \& {Blum}(2004)}]{Krause_Blum_2004_fractal}
{Krause}, M. \& {Blum}, J. 2004, \prl, 93, 021103

\bibitem[{{Krijt} {et~al.}(2014){Krijt}, {Dominik}, \&
  {Tielens}}]{Krijt_2014_rolling}
{Krijt}, S., {Dominik}, C., \& {Tielens}, A.~G.~G.~M. 2014, Journal of Physics
  D Applied Physics, 47, 175302

\bibitem[{Kruss {et~al.}(2016)Kruss, Demirci, Koester, Kelling, \&
  Wurm}]{kruss_failed_2016}
Kruss, M., Demirci, T., Koester, M., Kelling, T., \& Wurm, G. 2016, The
  Astrophysical Journal, 827, 110, aDS Bibcode: 2016ApJ...827..110K

\bibitem[{Kruss {et~al.}(2017)Kruss, Teiser, \& Wurm}]{kruss_growing_2017}
Kruss, M., Teiser, J., \& Wurm, G. 2017, Astronomy and Astrophysics, 600, A103

\bibitem[{Lambrechts \& Johansen(2012)}]{2012A&A...544A..32L}
Lambrechts, M. \& Johansen, A. 2012, Astronomy and Astrophysics, 544, 32

\bibitem[{{Lambrechts} {et~al.}(2019){Lambrechts}, {Morbidelli}, {Jacobson},
  {Johansen}, {Bitsch}, {Izidoro}, \& {Raymond}}]{2019A&A...627A..83L}
{Lambrechts}, M., {Morbidelli}, A., {Jacobson}, S.~A., {et~al.} 2019, \aap,
  627, A83

\bibitem[{Li \& Youdin(2021)}]{li_thresholds_2021}
Li, R. \& Youdin, A.~N. 2021, The Astrophysical Journal, 919, 107, aDS Bibcode:
  2021ApJ...919..107L

\bibitem[{{Mathis} {et~al.}(1977){Mathis}, {Rumpl}, \&
  {Nordsieck}}]{1977ApJ...217..425M}
{Mathis}, J.~S., {Rumpl}, W., \& {Nordsieck}, K.~H. 1977, \apj, 217, 425

\bibitem[{{Meeus} {et~al.}(2001){Meeus}, {Waters}, {Bouwman}, {van den Ancker},
  {Waelkens}, \& {Malfait}}]{2001A&A...365..476M}
{Meeus}, G., {Waters}, L.~B.~F.~M., {Bouwman}, J., {et~al.} 2001, \aap, 365,
  476

\bibitem[{{Min} {et~al.}(2005){Min}, {Hovenier}, \& {de Koter}}]{Min-DHS}
{Min}, M., {Hovenier}, J.~W., \& {de Koter}, A. 2005, \aap, 432, 909

\bibitem[{{Morris} {et~al.}(2012){Morris}, {Boley}, {Desch}, \&
  {Athanassiadou}}]{2012ApJ...752...27M}
{Morris}, M.~A., {Boley}, A.~C., {Desch}, S.~J., \& {Athanassiadou}, T. 2012,
  \apj, 752, 27

\bibitem[{Okuzumi(2009)}]{okuzumi_electric_2009}
Okuzumi, S. 2009, The Astrophysical Journal, 698, 1122, tex.ids:
  okuzumiElectricChargingDust2009b

\bibitem[{Okuzumi {et~al.}(2011{\natexlab{a}})Okuzumi, {Tanaka}, {Takeuchi}, \&
  Sakagami}]{okuzumi_electrostatic_2011-1}
Okuzumi, S., {Tanaka}, {Takeuchi}, \& Sakagami, M. 2011{\natexlab{a}}, The
  Astrophysical Journal, 731, 95

\bibitem[{Okuzumi {et~al.}(2011{\natexlab{b}})Okuzumi, {Tanaka}, {Takeuchi}, \&
  Sakagami}]{okuzumi_electrostatic_2011}
Okuzumi, S., {Tanaka}, {Takeuchi}, \& Sakagami, M. 2011{\natexlab{b}}, The
  Astrophysical Journal, 731, 96

\bibitem[{{Okuzumi} {et~al.}(2011){Okuzumi}, {Tanaka}, {Takeuchi}, \&
  {Sakagami}}]{2011ApJ...731...95O}
{Okuzumi}, S., {Tanaka}, H., {Takeuchi}, T., \& {Sakagami}, M.-a. 2011, \apj,
  731, 95

\bibitem[{Ormel \& Cuzzi(2007)}]{2007A&A...466..413O}
Ormel, C.~W. \& Cuzzi, J.~N. 2007, Astronomy and Astrophysics, 466, 413

\bibitem[{Ormel \& Klahr(2010)}]{2010A&A...520A..43O}
Ormel, C.~W. \& Klahr, H.~H. 2010, Astronomy and Astrophysics, 520, 43

\bibitem[{Paardekooper {et~al.}(2020)Paardekooper, McNally, \&
  Lovascio}]{paardekooper_polydisperse_2020}
Paardekooper, S.-J., McNally, C.~P., \& Lovascio, F. 2020, Monthly Notices of
  the Royal Astronomical Society, 499, 4223, aDS Bibcode: 2020MNRAS.499.4223P

\bibitem[{{Paardekooper} \& {Mellema}(2004)}]{2004A&A...425L...9P}
{Paardekooper}, S.~J. \& {Mellema}, G. 2004, \aap, 425, L9

\bibitem[{{Paardekooper} \& {Mellema}(2006)}]{2006A&A...453.1129P}
{Paardekooper}, S.~J. \& {Mellema}, G. 2006, \aap, 453, 1129

\bibitem[{Paszun \& {Dominik}(2009)}]{paszun_collisional_2009}
Paszun, D. \& {Dominik}. 2009, Astronomy and Astrophysics, 507, 1023

\bibitem[{Pinilla {et~al.}(2012)Pinilla, Birnstiel, Ricci, Dullemond, Uribe,
  Testi, \& Natta}]{pinilla_trapping_2012}
Pinilla, P., Birnstiel, T., Ricci, L., {et~al.} 2012, Astronomy \&amp;
  Astrophysics, Volume 538, id.A114,
  {\textless}NUMPAGES{\textgreater}15{\textless}/NUMPAGES{\textgreater} pp.,
  538, A114

\bibitem[{{Pinilla} {et~al.}(2012){Pinilla}, {Birnstiel}, {Ricci}, {Dullemond},
  {Uribe}, {Testi}, \& {Natta}}]{2012A&A...538A.114P}
{Pinilla}, P., {Birnstiel}, T., {Ricci}, L., {et~al.} 2012, \aap, 538, A114

\bibitem[{{Pinte} {et~al.}(2016){Pinte}, {Dent}, {M{\'e}nard}, {Hales}, {Hill},
  {Cortes}, \& {de Gregorio-Monsalvo}}]{2016ApJ...816...25P}
{Pinte}, C., {Dent}, W.~R.~F., {M{\'e}nard}, F., {et~al.} 2016, \apj, 816, 25

\bibitem[{{Rice}(2022)}]{Rice_2022_gravitational_collapse}
{Rice}, K. 2022, in Oxford Research Encyclopedia of Planetary Science (Oxford
  University Press), 250

\bibitem[{{Rice} {et~al.}(2006){Rice}, {Armitage}, {Wood}, \&
  {Lodato}}]{2006MNRAS.373.1619R}
{Rice}, W.~K.~M., {Armitage}, P.~J., {Wood}, K., \& {Lodato}, G. 2006, \mnras,
  373, 1619

\bibitem[{Schräpler {et~al.}(2012)Schräpler, Blum, Seizinger, \&
  Kley}]{schrapler_physics_2012}
Schräpler, R., Blum, J., Seizinger, A., \& Kley, W. 2012, The Astrophysical
  Journal, 758, 35, aDS Bibcode: 2012ApJ...758...35S

\bibitem[{Seizinger \& Kley(2013)}]{seizinger_bouncing_2013}
Seizinger, A. \& Kley, W. 2013, Astronomy and Astrophysics, 551, A65

\bibitem[{Sirono \& Ueno(2017)}]{sirono_collisions_2017}
Sirono, S.-i. \& Ueno, H. 2017, The Astrophysical Journal, 841, 36, aDS
  Bibcode: 2017ApJ...841...36S

\bibitem[{Stammler \& Birnstiel(2022)}]{stammler_dustpy_2022}
Stammler, S.~M. \& Birnstiel, T. 2022, The Astrophysical Journal, 935, 35, aDS
  Bibcode: 2022ApJ...935...35S

\bibitem[{{Stammler} \& {Birnstiel}(2022)}]{2022ApJ...935...35S}
{Stammler}, S.~M. \& {Birnstiel}, T. 2022, \apj, 935, 35

\bibitem[{{Stammler} {et~al.}(2023){Stammler}, {Lichtenberg},
  {Dr{\k{a}}{\.z}kowska}, \& {Birnstiel}}]{2023A&A...670L...5S}
{Stammler}, S.~M., {Lichtenberg}, T., {Dr{\k{a}}{\.z}kowska}, J., \&
  {Birnstiel}, T. 2023, \aap, 670, L5

\bibitem[{{Tazaki} {et~al.}(2023){Tazaki}, {Ginski}, \&
  {Dominik}}]{2023ApJ...944L..43T}
{Tazaki}, R., {Ginski}, C., \& {Dominik}, C. 2023, \apjl, 944, L43

\bibitem[{{Teague} {et~al.}(2016){Teague}, {Guilloteau}, {Semenov}, {Henning},
  {Dutrey}, {Pi{\'e}tu}, {Birnstiel}, {Chapillon}, {Hollenbach}, \&
  {Gorti}}]{2016A&A...592A..49T}
{Teague}, R., {Guilloteau}, S., {Semenov}, D., {et~al.} 2016, \aap, 592, A49

\bibitem[{{Testi} {et~al.}(2014){Testi}, {Birnstiel}, {Ricci}, {Andrews},
  {Blum}, {Carpenter}, {Dominik}, {Isella}, {Natta}, {Williams}, \&
  {Wilner}}]{2014prpl.conf..339T}
{Testi}, L., {Birnstiel}, T., {Ricci}, L., {et~al.} 2014, in Protostars and
  Planets VI, ed. H.~{Beuther}, R.~S. {Klessen}, C.~P. {Dullemond}, \&
  T.~{Henning}, 339--361

\bibitem[{{Villenave} {et~al.}(2022){Villenave}, {Stapelfeldt}, {Duch{\^e}ne},
  {M{\'e}nard}, {Lambrechts}, {Sierra}, {Flores}, {Dent}, {Wolff}, {Ribas},
  {Benisty}, {Cuello}, \& {Pinte}}]{2022ApJ...930...11V}
{Villenave}, M., {Stapelfeldt}, K.~R., {Duch{\^e}ne}, G., {et~al.} 2022, \apj,
  930, 11

\bibitem[{{Visser} \& {Ormel}(2016)}]{2016A&A...586A..66V}
{Visser}, R.~G. \& {Ormel}, C.~W. 2016, \aap, 586, A66

\bibitem[{Wada {et~al.}(2008)Wada, Tanaka, Suyama, Kimura, \&
  Yamamoto}]{wada_numerical_2008}
Wada, K., Tanaka, H., Suyama, T., Kimura, H., \& Yamamoto, T. 2008, The
  Astrophysical Journal, 677, 1296

\bibitem[{Wada {et~al.}(2011)Wada, Tanaka, Suyama, Kimura, \&
  Yamamoto}]{wada_rebound_2011}
Wada, K., Tanaka, H., Suyama, T., Kimura, H., \& Yamamoto, T. 2011, The
  Astrophysical Journal, 737, 36, aDS Bibcode: 2011ApJ...737...36W

\bibitem[{{Weidenschilling}(1977)}]{1977MNRAS.180...57W}
{Weidenschilling}, S.~J. 1977, \mnras, 180, 57

\bibitem[{Weidling \& Blum(2015)}]{weidling_free_2015}
Weidling, R. \& Blum, J. 2015, Icarus, 253, 31, aDS Bibcode:
  2015Icar..253...31W

\bibitem[{{Whipple}(1972)}]{1972fpp..conf..211W}
{Whipple}, F.~L. 1972, in From Plasma to Planet, ed. A.~{Elvius}, 211

\bibitem[{Wilner {et~al.}(2005)Wilner, D'Alessio, Calvet, Claussen, \&
  Hartmann}]{wilner_toward_2005}
Wilner, D.~J., D'Alessio, P., Calvet, N., Claussen, M.~J., \& Hartmann, L.
  2005, The Astrophysical Journal, 626, L109, aDS Bibcode: 2005ApJ...626L.109W

\bibitem[{Windmark {et~al.}(2012{\natexlab{a}})Windmark, Birnstiel, Güttler,
  Blum, Dullemond, \& Henning}]{windmark_planetesimal_2012}
Windmark, F., Birnstiel, T., Güttler, C., {et~al.} 2012{\natexlab{a}},
  Astronomy and Astrophysics, 540, A73

\bibitem[{Windmark {et~al.}(2012{\natexlab{b}})Windmark, Birnstiel, Ormel, \&
  Dullemond}]{windmark_breaking_2012}
Windmark, F., Birnstiel, T., Ormel, C., \& Dullemond, C. 2012{\natexlab{b}},
  Astronomy \& Astrophysics, 544, L16

\bibitem[{Xiang {et~al.}(2020)Xiang, Matthews, Carballido, \&
  Hyde}]{xiang_detailed_2020}
Xiang, C., Matthews, L.~S., Carballido, A., \& Hyde, T.~W. 2020, The
  Astrophysical Journal, 897, 182, aDS Bibcode: 2020ApJ...897..182X

\bibitem[{Youdin(2004)}]{youdin_obstacles_2004}
Youdin, A.~N. 2004, in Star Formation in the Interstellar Medium: In Honor of
  David Hollenbach Place: eprint: arXiv:astro-ph/0311191 ADS Bibcode:
  2004ASPC..323..319Y, Vol. 323, 319

\bibitem[{{Zsom} {et~al.}(2010){Zsom}, {Ormel}, {G{\"u}ttler}, {Blum}, \&
  {Dullemond}}]{2010A&A...513A..57Z}
{Zsom}, A., {Ormel}, C.~W., {G{\"u}ttler}, C., {Blum}, J., \& {Dullemond},
  C.~P. 2010, \aap, 513, A57

\bibitem[{Zsom {et~al.}(2010)Zsom, Ormel, Güttler, Blum, \&
  Dullemond}]{zsom_outcome_2010}
Zsom, A., Ormel, C.~W., Güttler, C., Blum, J., \& Dullemond, C.~P. 2010,
  Astronomy and Astrophysics, 513, A57

\bibitem[{{Zubko} {et~al.}(1996){Zubko}, {Mennella}, {Colangeli}, \&
  {Bussoletti}}]{1996MNRAS.282.1321Z}
{Zubko}, V.~G., {Mennella}, V., {Colangeli}, L., \& {Bussoletti}, E. 1996,
  \mnras, 282, 1321

\end{thebibliography}

\newpage
\begin{appendix} %First online appendix

\section{Diagnostics}
\label{app-diagnostics}
To analyze the results from the DustPy code, there were a number of
aposteriori (i.e., after the model run) diagnostics we calculated. In
this appendix, we give the definitions of these diagnostic quantities.

\subsection{Mean and width of the size distribution}
\label{app-mean-and-width}
At each radius $r$, we computed the mean grain size
\begin{equation}\label{eq-mean-a}
\langle a\rangle(r) = \frac{1}{\Sigma_{\mathrm{d}}(r)}\int_{m_{\mathrm{min}}}^{m_{\mathrm{max}}} \Sigma_{\mathrm{d}}(r,m) a(m) dm,
\end{equation}
where $\Sigma_{\mathrm{d}}(r)$ is the total dust surface density
\begin{equation}\label{eq-sigma-dust-total}
\Sigma_{\mathrm{d}}^{\mathrm{tot}}(r) = \int_{m_{\mathrm{min}}}^{m_{\mathrm{max}}} \Sigma_{\mathrm{d}}(r,m) dm,
\end{equation}
and $a(m)$ is
\begin{equation}\label{eq-a-afo-m}
a(m) = \left(\frac{3m}{4\pi\rho_{\bullet}}\right)^{1/3}.
\end{equation}

{Furthermore, we computed an indication a$_{\mathrm{max}}$ of the
  maximum grains size as the largest grain size where the
  $\Sigma_{\mathrm{d}}$ drops below $10^{-3}$ times the maximum value
  of $\Sigma_{\mathrm{d}}$ at that radius.}

We also note that in the DustPy code, the \texttt{Sigma} of the dust is
defined such that it is already multiplied by the mass bin width,
\texttt{dust.Sigma[i,n]}$\equiv \Sigma_{\mathrm{d}}(r_i,m_n)\Delta m_n$, so that
the integral of Eq.~(\ref{eq-sigma-dust-total}) is, in Python, just
the sum \texttt{dust.Sigma.sum(axis=-1)}.

The width of the distribution was computed as the standard deviation
$\sigma^2_{\ln(a)}$ of $\ln(a)$ from the value
$\ln(\langle a\rangle(r))$:
\begin{equation}\label{eq-sigma-ln-a}
\sigma^2_{\ln(a)} = \frac{1}{\Sigma_{\mathrm{d}}(r)}\int_{m_{\mathrm{min}}}^{m_{\mathrm{max}}} \Sigma_{\mathrm{d}}(r,m) [\ln(a(m))-\ln(\langle a\rangle(r))]^2 dm.
\end{equation}
The reason we defined the width in natural-log space instead of in linear space
is that for the models without the bouncing barrier, the size distribution
becomes rather wide, spanning several orders of magnitude in grain radius $a$,
which is difficult to capture in linear $a$ space. We used the natural log
instead of the 10-log because in the limit of small values of the
standard deviation $\sigma^2_{\ln(a)}$, this value becomes equal to
the linear standard deviation divided by the mean squared,
\begin{equation}
\lim_{\sigma^2\rightarrow 0} \sigma^2_{\ln(a)} = \frac{\sigma^2_{a}}{\langle a\rangle^2},
\end{equation}
which can be proven by employing a Taylor expansion. Hence, for $\sigma^2_{\ln(a)}\ll 1$,
we can unambiguously state that the size distribution is narrow, while for
$\sigma^2_{\ln(a)}\gtrsim 1$, it is wide.

\subsection{Optical depth of the dust distribution}
\label{sec:optical-depth}
To compute the optical depth, we first needed to compute the opacities of
the different grain sizes. We used \texttt{optool}
\citep{2021ascl.soft04010D} with the DHS (distribution of hollow
spheres) formalism \citep{Min-DHS}, setting the $f_{\mathrm{max}}$
parameter to 0.8. The mixture was 87\% pyroxene, with 70\% magnesium
\citep{1995A&A...300..503D}, and the other 13\% was amorphous carbon
\citep{1996MNRAS.282.1321Z}. The porosity was set to 0.397 so that the
mean material density of the dust aggregates would be $\rho_{\bullet}=1.67$,
consistent with the \texttt{DustPy} model. With these opacities, we could then
compute the total vertical optical depth at wavelength $\lambda$ at
location $r$:
\begin{equation}
\tau_\lambda(r) = \int_{m_{\mathrm{min}}}^{m_{\mathrm{max}}}\Sigma_{\mathrm{d}}(r,m)\kappa_\nu(m)\,dm.
\end{equation}

\subsection{Images of the disk}
To make images of the disk that could be compared to observations, we used the
RADMC-3D radiative transfer code \citep{2012ascl.soft02015D}. We made one
image at $\lambda=1\mu$m at an inclination $i=70^{\circ}$ and another image
at $\lambda=1.3$mm at an inclination $i=90^{\circ}$. By choosing $i=70^{\circ}$ for
the $1\mu$m image, we could better see the shape of the surface layers, while
by choosing $i=90^{\circ}$ for the 1.3 mm image, we could better see the vertical
extent of the dust.

\subsection{Degree of settling}
DustPy stores the vertical extent $H_{\mathrm{d}}(r,m)$ of each grain mass and at
each location. To obtain the mean dust vertical extent $H_{\mathrm{d}}(r)$ we
computed
\begin{equation}
H^{\mathrm{tot}}_{\mathrm{d}}(r) = \frac{1}{\Sigma_{\mathrm{d}}(r)}\int_{m_{\mathrm{min}}}^{m_{\mathrm{max}}} H_{\mathrm{d}}(r,m)\Sigma_{\mathrm{d}}(r,m)dm.
\end{equation}

\subsection{Number of collisions per megayear}
The bouncing barrier is thought to be accompanied or even triggered by
collisional compaction \citep{2010A&A...513A..57Z}. Dust aggregates
tend to initially grow in a "fluffy" manner, often called "fractal
growth." That is because each collision may lead to some degree of
compactification but also to growth. At some point, the combination
of slight compaction and increased velocities are then assumed to lead
to the appearance of bouncing. From there, collisions no longer
lead to growth, and the effect of further collisions can then
successively compactify the aggregates.  These aggregates remain
porous, but they lose their fractal "fluffy" shape.  Being more
round and compact to some degree, they tend to bounce even more
(i.e., stick even less) to similar-sized aggregates. This reinforces the
bouncing barrier, as is shown very clearly in Figure 4b of
\citet{2010A&A...513A..57Z}, where the "enlargement factor" is
reduced during bouncing, and Figure 4c, where the "bouncing with
compaction" dominates after the bouncing barrier has been
reached. The compaction, however, requires some time, or better, it
requires numerous successive collisions. Exact quantitative values of
the required number of collisions are not known, but it is worthwhile
to find out whether in the current model, the expected number of
collisions is much larger than unity or not.  Of course, our model
does not include the effect of fractal growth, but if it did, it
would likely increase the collision rate rather than decrease it, so
our model can provide a lower limit to the number of collisions.
If the result is that the number of collisions is much larger than unity, then
we may expect the dust aggregates to have time to become nicely rounded-off
and "pebble-like."

The simplest definition of the mean collision rate (the number of collisions per
dust aggregate per second) is
\begin{equation}\label{eq-Rcoll-all}
  R_{\mathrm{coll}}^{\mathrm{tot}}(m) = \int_{m_{\mathrm{min}}}^{m_{\mathrm{max}}}
  \sigma_{\mathrm{coll}}(m,m')\,\Delta v(m,m')\,n(m')\,dm',
\end{equation}
where $\sigma_{\mathrm{coll}}(m,m')=\pi(a+a')^2$ (using
Eq.~\ref{eq-a-afo-m}) is the collisional cross section,
$\Delta v(r,m,m')$ is the expected relative collisional velocity, and
$n(r,m')=\Sigma_{\mathrm{d}}(r,m')/(\sqrt{2\pi}h_{\mathrm{d}}(r,m')m')$
is the midplane volume number density of particles with mass $m'$,
where $h_{\mathrm{d}}(r,m')$ is the vertical extent of the layer of
these dust aggregates.

The problem with this definition of the number of collisions is that it includes
collisions of a dust aggregate with any mass, including tiny (micron sized) dust
particles. This means that for a millimeter size dust aggregate, it also includes
collisions with micron-sized particles. This may lead to very large numbers, even
if the number of collisions between similar-sized aggregates remains low. A
better definition that accounts for the mass ratio of the particles is
\begin{equation}\label{eq-Rcoll-massratio}
  R_{\mathrm{coll}}^{\mathrm{ratio}}(r,m) = \int\limits_{m_{\mathrm{min}}}^{m_{\mathrm{max}}}
  \sigma_{\mathrm{coll}}(m,m')\,\Delta v(r,m,m')\,\min\left(\frac{m'}{m},1\right)\,n(r,m')\,dm',
\end{equation}
where we scale the weight of collisions with very small collision
partners with the mass ratio. Alternatively, we could simply \rev{limit
ourselves to}
collisions with almost equal sized dust aggregates:
\begin{equation}\label{eq-Rcoll-almostequal}
  R_{\mathrm{coll}}^{\mathrm{equ}}(r,m) = \int_{m/2}^{2m}
  \sigma_{\mathrm{coll}}(r,m,m')\,\Delta v(r,m,m')\,n(r,m')\,dm'.
\end{equation}

Next, for each of these definitions, we took the mean over mass:
\begin{equation}\label{eq-mean-coll-rate}
\bar R_{\mathrm{coll}}^{\mathrm{X}}(r) = \frac{1}{\Sigma^{\mathrm{tot}}_{\mathrm{d}}(r)}\int_{m_{\mathrm{min}}}^{m_{\mathrm{max}}} \Sigma_{\mathrm{d}}(r,m)\,R_{\mathrm{coll}}^{\mathrm{X}}(r,m)\,dm,
\end{equation}
where X is a placeholder for tot, ratio, or equ.  We re-inserted the
$r$-dependency here for clarity.

Finally, to use a more meaningful timescale, we estimated the number
of collisions per megayear simply by multiplying by
$3.15576\times 10^{13}$ seconds:
\begin{equation}
\bar R_{\mathrm{coll,Myr}}^{\mathrm{X}}(r) = 3.15576\times 10^{13}\;\bar R_{\mathrm{coll}}^{\mathrm{X}}(r).
\end{equation}

\subsection{Mean collisional velocity}
The mean collisional velocity was computed for "almost equal size" collisions,
as in Eq.~(\ref{eq-Rcoll-almostequal}). The integral becomes quadratic in
$\Delta v(r,m,m')$ because we need one factor $\Delta v$ for the collision rate
and another factor $\Delta v$ for computing the mean:
\begin{equation}\label{eq-vcoll-almostequal}
\begin{split}
  \langle\Delta v_{\mathrm{coll}}^{\mathrm{equ}}(r)\rangle = & \bar R_{\mathrm{coll}}^{\mathrm{X}}(r)^{-1}\, \int_{m/2}^{2m}\int_{m/2}^{2m} 
  \sigma_{\mathrm{coll}}(r,m,m')\; \times \\
  & \,\Delta v(r,m,m')^2\,n(r,m')\,n(r,m)\,dm'dm.
\end{split}
\end{equation}

\section{Size distributions for the turbulence series and miscellaneous series}\label{app-series-sigma}
In Sections \ref{sec-series-alpha} and \ref{sec-series-misc}, the results for
the two series of models were discussed. Here we show the full distribution function
data.

\begin{figure*}
  \includegraphics[width=\textwidth]{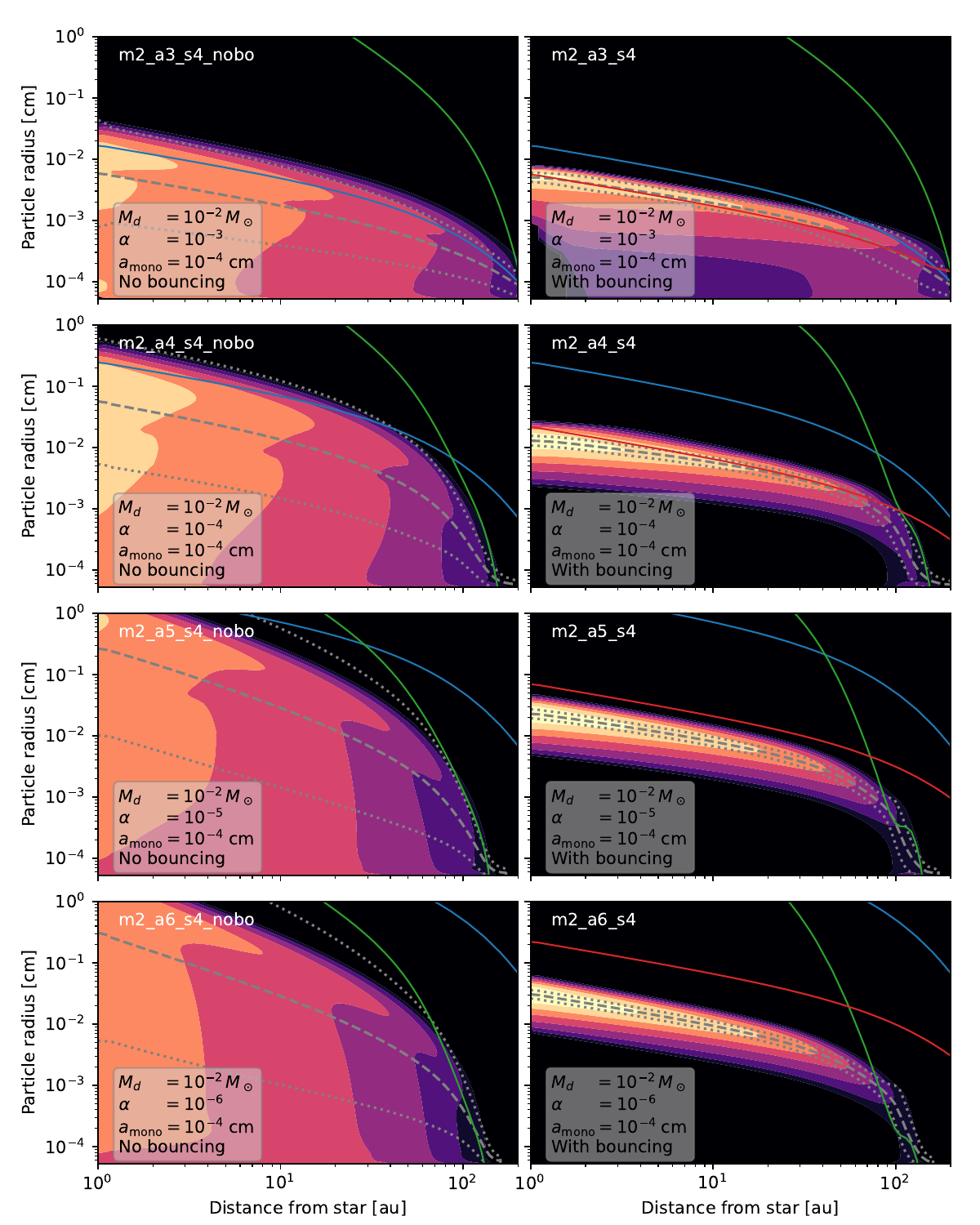}
  \caption{\label{fig-series-alpha-sigmadust}Dust surface densities for the
    turbulence series (Section \ref{sec-series-alpha}). Left: Turbulence series without bouncing. Right: Turbulence series with
    bouncing.
  }
\end{figure*}

\begin{figure*}
  \includegraphics[width=1.03\textwidth]{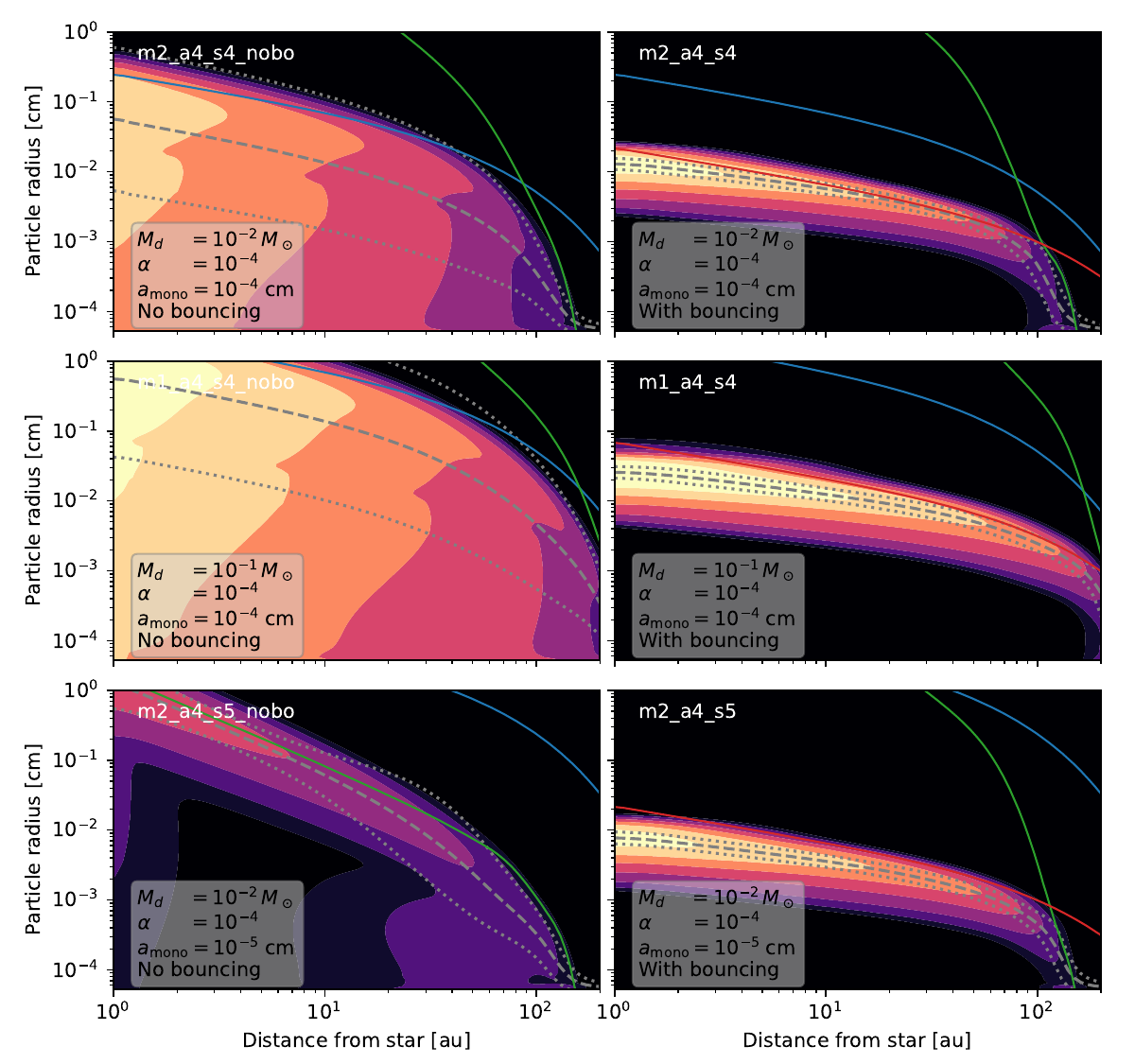}
  \caption{\label{fig-series-mixed-sigmadust}Dust surface densities for the
    miscellaneous series (Section \ref{sec-series-misc}). Left: Miscellaneous series without bouncing. Right: miscellaneous series with
    bouncing.
  }
\end{figure*}
\FloatBarrier

\section{Resolution test}\label{app-resolution}
\begin{figure}[t!]
  \includegraphics[width=.45\textwidth]{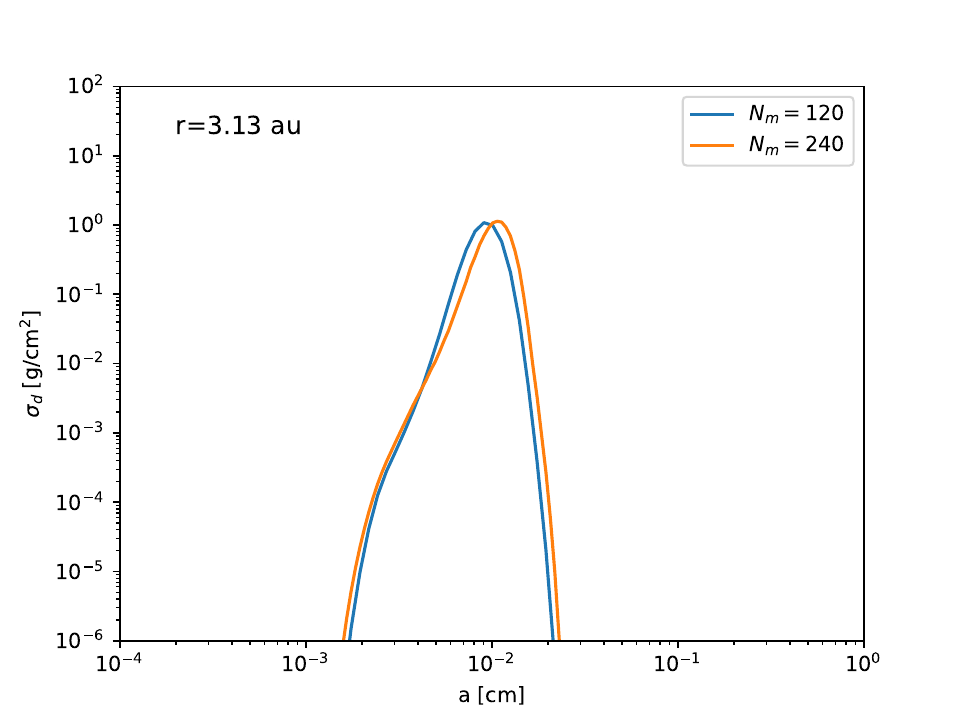}
  \caption{\label{fig-resolution-slice-sigma}As in Fig.~\ref{fig-fidu-slice-sigma} but now
    the model with the fiducial mass resolution ($N_m=120$) is compared to a run with
    double the resolution ($N_m=240$).}
\end{figure}

To ascertain that the mass resolution is sufficient to capture the
narrow mass distributions we obtained with the bouncing barrier, we redid
the fiducial model with twice the number of mass sampling points
($N_m=240$) and compared the distributions at $r=3.13$ au. As shown in
Fig.~\ref{fig-resolution-slice-sigma}, the result shows that while the
distribution peak in the fiducial resolution model is slightly shifted
with respect to the high resolution model, the overall match is
satisfactory.

\begin{figure}[t!]
  \includegraphics[width=.45\textwidth]{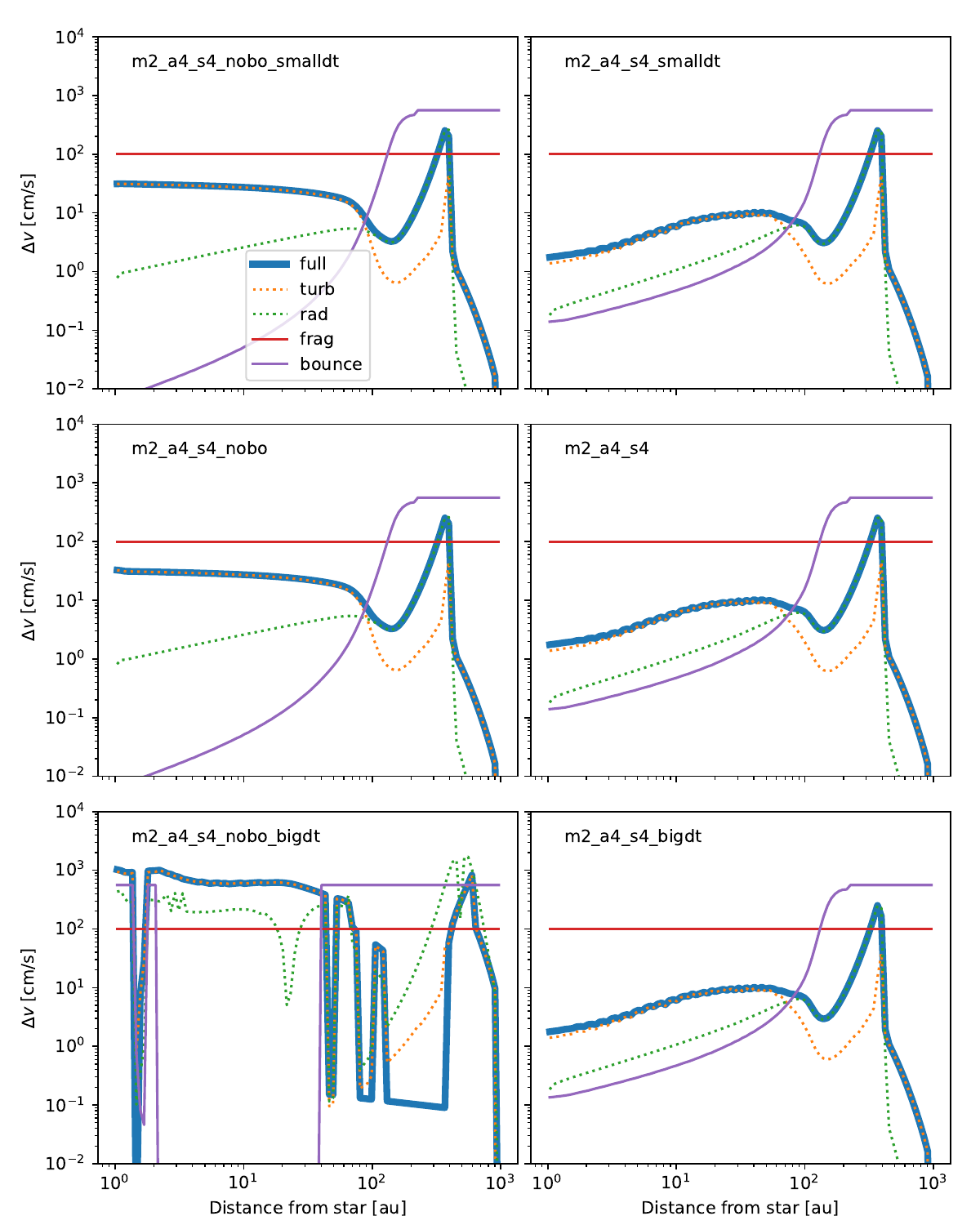}
  \caption{\label{fig-time-resolution-vcoll}As in
    Fig.~\ref{fig-fidu-collision-velocities} but created from
    runs with different time steps with boosting factors 1, 10, and
    100, from top to bottom.}
\end{figure}

To ascertain the correct choice of time steps for the integration, we
made runs with different time step boost factors.  The effects of
this can be seen in many model outcomes. We chose the plot
showing the collision velocities.  The results are shown in
fig.~\ref{fig-time-resolution-vcoll} and demonstrate that the results
are very similar with boost factors of 1 and 10, but that the computation
becomes unstable with a boost factor of 100.

\FloatBarrier
\vspace*{\fill}

\section{Code snippets for implementing the bouncing barrier into
  DustPy}
\label{app-code-snippets}
To implement the bouncing barrier into DustPy, we first (before
calling \texttt{sim.run()}) computed the bouncing velocity $\vbounce$ 
\begin{lstlisting}
sim.dust.froll    = 1e-4       # Heim et al. 1999
sim.dust.mredu    = sim.grid.m[:,None] * sim.grid.m[None,:] / ( sim.grid.m[:,None] + sim.grid.m[None,:] )
sim.dust.v.bounce = np.sqrt(5*np.pi*sim.dust.amono* sim.dust.froll/sim.dust.mredu) # Eq. 7 of Guettler et al. 2010
\end{lstlisting}
Next we overrode the \texttt{sim.dust.p.stick()} function of DustPy by
defining our own
\begin{lstlisting}
def dd_p_stick(sim):
    dum               = (sim.dust.v.bounce[None,:,:]/ sim.dust.v.rel.tot)**2
    sim.dust.p.bounce = (1.5*dum + 1) * np.exp(-1.5*dum)
    pnostick          = np.maximum(sim.dust.p.frag, sim.dust.p.bounce)
    p                 = 1. - pnostick
    p[0]              = 0.
    p[-1]             = 0.
    return p
\end{lstlisting}
and we linked it in through the \rev{following assignment.}
\begin{lstlisting}
sim.dust.p.stick.updater = dd_p_stick
\end{lstlisting}

\section{Table of symbols}

Here is a  table of often-used symbols of this
paper. The dimension column gives the dimension of the quantity. The
last column gives the equation or section where the symbol first
appears.\\[4mm]
\begin{tabular}{@{}llll}
    Symbol                               & Meaning                                   & Dim    & Eq/Sec                            \\
    \hline
    $t$                                  & Time                                      & s            &                           \\
    $\sigma_{\mathrm{H}_2}$              & Cross section of (H$_2$) collisions   & cm$^2$       & Sec.~\ref{sec:estimate-bb}        \\
    $\sigma_{\mathrm{coll}}$             & Cross section dust prtcl collisions & cm$^2$       & Eq.~\ref{eq-Rcoll-all}            \\
    $\sigma_{\ln(a)}$                    & Width of size distribution                & -            & Eq.~\ref{eq-sigma-ln-a}           \\
    $\sigmagas$                          & Gas surface density                       & g\,cm$^{-2}$ & Eq.~\ref{eq-gas-surface-density}  \\
    $\Sigma_{\mathrm{d}}^{\mathrm{tot}}$ & Totad dust surface density                & g\,cm$^{-2}$ & Eq.~\ref{eq-sigma-dust-total}     \\
    $\St$                                & Stokes number grain size $a$              & -            & Eq.~\ref{eq:stokes-number}        \\
    $\St_{\mathrm{b0}}$                  & St at bouncing  barrier        & -            & Eq.~\ref{eq-stokes-bounce}      \\
    $\St_{\mathrm{f}}$                   & St at fragmentation barrier    & -            & Eq.~\ref{eq:stokes-fragmentation} \\
    $\abouncezero$                       & Size limit due to bouncing                & cm           & Eq.~\ref{eq:abouncezero}          \\
    $\vstick$                            & Sticking velocity                         & cm\,s$^{-1}$ & Eq.~\ref{eq:vstick}               \\
    $\vbounce$                           & Bouncing velocity                         & cm\,s$^{-1}$ & Eq.~\ref{eq-vbounce}              \\
    $\vfrag$                            & Fragmentation velocity                     & cm\,s$^{-1}$ & Eq.~\ref{eq:vfragscale}           \\
    $\sigma^2_{\ln(a)}$                  & Width of size distribution peak       & -            & Eq.~\ref{eq-sigma-ln-a}           \\
    $\langle a\rangle$                   & Mean grain size                           & cm           & Eq.\ref{eq-mean-a}                \\
\end{tabular}

\end{appendix}

\end{document}